\documentclass[10pt,journal,compsoc]{IEEEtran}

\usepackage{ifpdf}
\usepackage{xspace}
\usepackage[table]{xcolor}
\usepackage{graphicx}
\usepackage{caption}
\usepackage{subcaption}
\usepackage{cite}
\usepackage{todonotes}
\usepackage{afterpage}
\usepackage[inline]{enumitem}
\usepackage{float}
\usepackage{multitoc}
\usepackage{multirow}
\usepackage{booktabs}
\usepackage{tcolorbox}
\usepackage{amsmath,amssymb}
\usepackage{url}
\usepackage{cleveref}
\usepackage{dblfloatfix}
\usepackage[numbers]{natbib}

\definecolor{gray50}{gray}{.5}
\definecolor{gray40}{gray}{.6}
\definecolor{gray30}{gray}{.7}
\definecolor{gray20}{gray}{.8}
\definecolor{gray10}{gray}{.9}
\definecolor{gray05}{gray}{.95}

\newlength\Linewidth
\def\findlength{\setlength\Linewidth\linewidth
	\addtolength\Linewidth{-4\fboxrule}
	\addtolength\Linewidth{-3\fboxsep}
}
\newenvironment{rqbox}{\par\begingroup
	\setlength{\fboxsep}{5pt}\findlength
	\setbox0=\vbox\bgroup\noindent
	\hsize=0.95\linewidth
	\begin{minipage}{0.95\linewidth}\normalsize}
	{\end{minipage}\egroup
	\textcolor{gray20}{\fboxsep1.5pt\fbox
		{\fboxsep5pt\colorbox{gray05}{\normalcolor\box0}}}
	\endgroup\par\noindent
	\normalcolor\ignorespacesafterend}

\newcounter{Finding}
\stepcounter{Finding}

\newcommand{\etal}{\textit{et al}. }
\newcommand{\ie}{\textit{i}.\textit{e}.,\xspace}
\newcommand{\eg}{\textit{e}.\textit{g}.,\xspace}

\newcommand{\code}[1]{\texttt{#1}}

\newcommand{\dataset}{Commit-Set based dataset\xspace}
\newcommand{\issueTracker}{Bugzilla\xspace}
\newcommand{\TheIssueTracker}{Bugzilla\xspace}
\newcommand{\MozillaLong}{Mozilla\xspace}

\newcommand{\numDevelopers}{608\xspace}
\newcommand{\BugzillaReports}{Bugzilla reports\xspace} 

\newcommand{\mozilla}{Mozilla\xspace}
\newcommand{\Mozilla}{Mozilla\xspace}
\newcommand{\numOfLinkedPRs}{5,348\xspace}
\newcommand{\numOfLinkedCommits}{24,089\xspace}
\newcommand{\numOfTotalPRs}{9,110\xspace}
\newcommand{\numOfLinkingDevs}{608\xspace}
\newcommand{\linkingDuration}{22\xspace}

\newcommand{\szz}{\textsc{szz}\xspace}
\newcommand{\bszz}{\textsc{b-szz}\xspace}
\newcommand{\agszz}{\textsc{ag-szz}\xspace}
\newcommand{\lszz}{\textsc{l-szz}\xspace}
\newcommand{\rszz}{\textsc{r-szz}\xspace}
\newcommand{\maszz}{\textsc{ma-szz}\xspace}
\newcommand{\raszz}{\textsc{ra-szz}\xspace}
\newcommand{\pydriller}{\textsc{PyDriller}\xspace}

\newcommand{\rosab}{Rosa's Benchmark\xspace}
\newcommand{\mozillab}{Mozilla's Benchmark\xspace}
\newcommand{\bugin}{fix-inducing\xspace}
\newcommand{\bugout}{bug-fixing\xspace}

\newcommand{\singlevarLabeling}{Single-variation labeling\xspace}
\newcommand{\allvarLabeling}{All-variations labeling\xspace}
\newcommand{\singlevarBased}{Single-variation based\xspace}
\newcommand{\allvarBased}{All-variations based\xspace}

\newcommand{\feat}[1]{`\texttt{#1}'}

\newboolean{showcomments}
\setboolean{showcomments}{true}
	
\ifthenelse{\boolean{showcomments}}
{
	\newcommand{\nb}[3]{
		{\colorbox{#2}{\bfseries\sffamily\scriptsize\textcolor{white}{#1}}}
		{\textcolor{#2}{\textsf\small{#3}$\blacktriangleleft$}}}
	 
}{
	\newcommand{\nb}[3]{}
} 

\newcommand{\hide}[1]{}

\definecolor{purple}{HTML}{DADAEB}
\definecolor{blue1}{HTML}{e1effc}
\definecolor{palevioletred}{HTML}{db7093}

\definecolor{LightCyan}{rgb}{0.88,1,1}

\begin{document}

	\title{SZZ in the Time of Pull Requests}
	
	\author{Fernando~Petrulio,		
			David~Ackermann,
			Enrico~Fregnan,
			G\"{u}l~Calikli,
			Marco~Castelluccio,
			Sylvestre~Ledru,
			Calixte~Denizet,
			Emma~Humphries,
			and~Alberto~Bacchelli
		
		\IEEEcompsocitemizethanks{
			\IEEEcompsocthanksitem F. Petrulio, D. Ackermann, E. Fregnan, and A. Bacchelli are with ZEST in the Department of Informatics at the University of Zurich, Switzerland. \\
			E-mail: fpetrulio@ifi.uzh.ch
			\IEEEcompsocthanksitem G. Calikli is with the School of Computing Science, University of Glasgow, UK.
			\IEEEcompsocthanksitem M. Castelluccio, S. Ledru, C. Denizet, and E. Humpries are with Mozilla Corporation. 
		}

		\thanks{Manuscript received . . . ; revised . . .}
}
	
	\markboth{IEEE Transactions on Software Engineering}{}

	\IEEEtitleabstractindextext{%
		\begin{abstract}
		In the multi-commit development model, programmers complete tasks (\eg implementing a feature) by organizing their work in several commits and packaging them into a commit-set. 
		Analyzing data from developers using this model can be useful to tackle challenging developers' needs, such as knowing which features introduce a bug as well as assessing the risk of integrating certain features in a release.
		However, to do so one first needs to identify \emph{\bugin} commit-sets. For such an identification, the \szz algorithm is the most natural candidate, but its performance has not been evaluated in the multi-commit context yet.
		
		In this study, we conduct an in-depth investigation on the reliability and performance of \szz in the multi-commit model. To obtain a reliable ground truth, we consider an already existing \szz dataset and adapt it to the multi-commit context. Moreover, we devise a second dataset that is more extensive and directly created by developers as well as Quality Assurance (QA) engineers of Mozilla. Based on these datasets, we (1) test the performance of \bszz and its non-language-specific \szz variations in the context of the multi-commit model, (2) investigate the reasons behind their specific behavior, and (3) analyze the impact of non-relevant commits in a commit-set and automatically detect them before using \szz.
		\end{abstract}
		
		}

	\maketitle

	\IEEEdisplaynontitleabstractindextext

	\IEEEpeerreviewmaketitle

\section{Introduction}\label{prszz_sec:intro}

\IEEEPARstart{M}{any} software projects adopt the \emph{multi-commit development model}~\cite{gousios2014exploratory, barr2012cohesive, Gousios:2015}.
In this model, developers complete their tasks (\eg fixing a bug or implementing a new feature) by organizing their work into several commits packaged into a \emph{commit-set}.
Among the various instances of the multi-commit model, the pull-based development model proposed by GitHub is the most popular~\cite{gousios2014exploratory}. 

The availability of data accumulated by projects using the multi-commit model opens up new opportunities for research to understand and support developers' needs.
By using a multi-commit model developers can bundle all the commits they work on when implementing a \emph{feature} or \emph{bug fix} in a single commit-set.
These commit-sets can be used by researchers to have access to feature level information, which, in turn, can be used to conduct feature-level defect prediction, features' risk assessment, as well as empirical studies on software quality involving features.

Feature-level information is considered important by practitioners. For example, a recent study~\cite{wan2018perceptions} found that developers are interested in knowing which feature introduced a defect (rather than which commit, component, or even method as previously thought~\cite{kamei2016defect}). 
Furthermore, release engineers need to determine how risky each feature is~\cite{saliu2005supporting} when deciding which features to integrate into a release~\cite{adams2016modern}.
%

To use software evolution data (such as the multi-commit one) for most applications related to software quality, one needs to know \emph{which changes introduced a defect}~\cite{rodriguez2018reproducibility}. In 2005, {\'S}liwerski, Zimmermann, and Zeller devised an algorithmic approach (known as \szz) to detect the commit(s) responsible for introducing a specific defect in a software system~\cite{sliwerski2005changes}. Thanks to \szz, researchers could conduct impactful studies on software quality~\cite{rodriguez2018reproducibility}.

\szz is the natural candidate for detecting which commit-sets introduce defects in the multi-commit model.
However, \szz and its variations~\cite{kim2006automatic,davies2014comparing,neto2018impact,neto2019} work at the \emph{commit level}\footnote{Commit-level: the algorithm's input is a single fixing commit and its output is a set of one or more candidate \bugin commits.} and have been evaluated accordingly. Therefore, we do not know whether \szz can be used reliably in the multi-commit context and, for example, whether \szz variations retain their benefits and if one should adapt the input and output of \szz.

In this paper, we present an in-depth empirical evaluation of \szz in the multi-commit development model. We investigated how \szz and its variations perform in this context and how they can be adapted.

As the first step in our study, we focused on obtaining a solid benchmark on which to evaluate \szz in the multi-commit context. We tackled this from two angles. On the one hand, we considered the most recent and comprehensive benchmark created to evaluate \szz variations at the commit level~\cite{pascarella2021evaluating}. We adapted this benchmark for the multi-commit context. On the other hand, we teamed up with \mozilla to obtain a novel, developer-created dataset specific to the multi-commit model. In this case, we designed and deployed a new dedicated data field for the issue tracker tool used by \mozilla: \issueTracker. Through this new field practitioners can specify the \emph{commit-set} whose implementation induced the bug at hand. Developers, Quality Assurance (QA) engineers, users, and automated tools (overseen by developers) have been using this extension for more than \linkingDuration months. This effort resulted in the creation of our second dataset, which comprises manually validated links between \numOfLinkedPRs \emph{commit-set}s, for a total of \numOfLinkedCommits commits.

In the second step of our study, we evaluated the performance of \szz and its variations in the multi-commit context. The more extensive benchmark (\ie the \mozilla one) contains different languages and technologies, so we focused on non-language-specific \szz variations. To gain deeper knowledge on the factors that improve or reduce the performance of the algorithms, we also manually analyzed 262 cases split among the unique findings and mistakes of each \szz variation.

Finally, since \szz gets commits as input and not all the commits in a \emph{commit-set} may help \szz link the right \bugin \emph{commit-set}s, we investigated the impact of removing the non-relevant input for \szz. We also explored multiple machine learning models to automatically remove non-relevant input and evaluated their impact on \szz's results.

\section{Background and Related Work}\label{prszz_sec:background}

We provide background on \szz, its variations, and the multi-commit development model, particularly its pull-based form. We also describe empirical studies that used \szz and its variations, as well as empirical studies on the multi-commit model. We conclude motivating why it is important and timely to evaluate ways to identifying bug-inducing commit-sets, in addition to single bug-inducing commits.

 \begin{figure}[ht]
	\includegraphics[width=\linewidth]{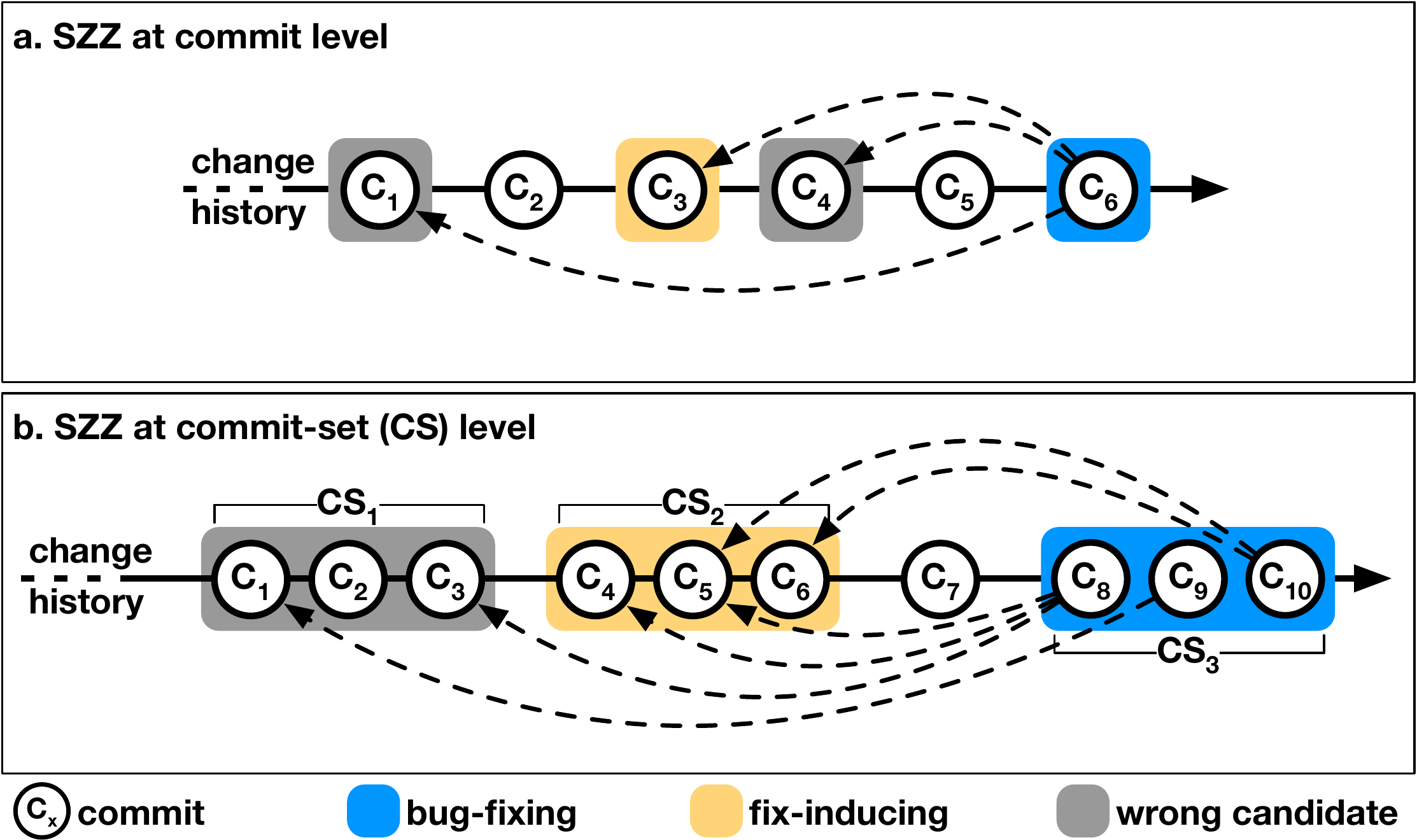}
	
	\caption{SZZ at (a) commit level and (b) commit-set level (\ie in the multi-commit development model)}
	\label{fig:szz_workflow}	
\end{figure}

\subsection{\bszz: The Original \szz Algorithm} \label{prszz_sec:back_szz_how}
The original version of \szz~\cite{sliwerski2005changes}, usually referred to as \bszz \cite{da2016framework, pascarella2021evaluating}, consists of two consecutive stages:
(1) identifying \bugout changes and
(2) identifying \bugin changes.

The first stage is concerned with linking bug reports to the commits fixing them. The difficulty of this step is dictated by the specific tools and development process used in the software systems under analysis.
Nowadays, the first stage can be completed reliably and with minor effort. Most contemporary software projects use issue tracking systems (\eg Bugzilla, GitHub) that maintain a link between each issue and the commits (or commit-sets) solving it.

The second stage's goal is to determine which commit(s) introduced a bug that was fixed by the fixing commit at hand. In this stage, the \bszz algorithm receives as input a \bugout commit (as identified in stage 1) and the history of commits. By using the \texttt{annotate} command of CVS, for each line in the bug fixing commit (depicted as $c_{6}$ in \Cref{fig:szz_workflow}-A), the algorithm finds the most recent commit (\eg $c_{4}$ in \Cref{fig:szz_workflow}-A) that modified the line before the fix. Since a commit can contain more than one line and each line might have been modified in different past commits, several commits can be marked as \bugin. In the example in \Cref{fig:szz_workflow}-A, the output of \bszz is the set of commits \{$c_1$, $c_3$, $c_4$\}.
Accurately completing the second stage is an open research challenge and is the focus of the work on \szz variations as well as our study.

\subsection{\szz Variations}\label{prszz_sec:back_szz_versions}
In the following, we focus on the \szz variations we consider in our investigation.
Our empirical evaluation takes inspiration from the study by Rosa \etal~\cite{pascarella2021evaluating}, which is the most recent and comprehensive study assessing \bszz and its variations at commit level. Therefore, for our assessment, we started by considering all the variations analyzed in the study~\cite{pascarella2021evaluating}.
Then, we excluded those that work only on Java files (\raszz$^\ast$~\cite{neto2019} and OpenSZZ~\cite{Lenarduzzi:2020}) or on only Java/Javascript/C files (\maszz~\cite{da2016framework}). In fact, in the \mozilla dataset we created for our study, Java files are only 0.34$\%$ of all files, and Javascript/C ones are less than 30\%. We also had to exclude \szz Unleashed~\cite{borg2019szz} because it could not scale up to the size of the source code repository of \mozilla.

\smallskip
\noindent \textbf{\agszz.}\label{prszz_sec:back_szz_v_ag}
\bszz uses the \code{annotate} command of CVS to track down modified/deleted lines. However, \code{annotate} cannot detect changes in methods' names. Therefore, whenever the name of a method containing a buggy line is changed, \bszz cannot do the mapping and \bugin changes are undetected (thus increasing false negatives). \agszz solves this issue by using \textit{annotation graphs}~\cite{kim2006automatic}.
\agszz also reduces false positives by ignoring format changes and changes to comments and blank lines. 

\smallskip
\noindent \textbf{\lszz \& \rszz.}\label{prszz_sec:back_szz_v_r}
The \agszz algorithm may return multiple \bugin commits for each \bugout commit. \lszz and \rszz employ heuristics to filter the output that \agszz produces to return only one \bugin commit for each \bugout commit. \lszz uses a heuristic to return the change with the highest number of added/deleted/changed lines, whereas \rszz returns the most recent commit among all the candidates.\smallskip 

\smallskip
\noindent \textbf{\pydriller.}\label{prszz_sec:back_szz_v_pydriller}
\pydriller~\cite{spadini2018pydriller}, a tool for mining software repositories, contains an enhanced implementation of \agszz. \pydriller uses \code{git-hyper-blame} to bypass meta changes that do not change software behavior (\eg refactorings) to decrease false positives. Such changes need to be manually specified in a dedicated file containing all revisions (commit hashes) to exclude from \szz computation.

\subsection{Applications of \szz in Empirical SE}
Most of the applications of \bszz and \szz variations in empirical SE research focus on software quality aspects.

\smallskip
\noindent \textbf{Defect prediction.} In the study where Herbold \etal~\cite{herbold2019issues} analyzed the issues with \bszz, the authors reported that as of August 2019, slightly more than 50\% (8 out of 15) of existing public datasets used for defect prediction research were labeled by using \bszz. Various studies in the literature prepared their own datasets (\ie ground truth) by using \bszz or its variations to learn and evaluate the defect prediction models~\cite{Toth:2016, pascarella2019, Kim:2008, Wen:2016, yan2020,kim2007predicting, Rahman:2011,chen2019extracting}. However, the accuracy of the defect prediction models relies on the accuracy of the \szz approach used for labeling the dataset~\cite{Fan:2019}. Fan \etal~\cite{Fan:2019} empirically investigated the effect of mislabeled data on defect prediction performance. The authors reported that mislabeled changes by \agszz lead to a statistically significant reduction in prediction accuracy; in contrast, data mislabeled by \bszz and \maszz do not cause a considerable reduction in prediction models' accuracy.

\noindent \textbf{Analysis of software quality related factors.}
Researchers used the datasets labeled by \bszz and \szz variations to investigate how software quality relates to several factors empirically~\cite{aman2019empirical, eyolfson2014correlations, bernardi2018relation, Izquierdo-Cortazar2012, tufano2017empirical, chen2019extracting}. 
For instance, Eyolfson \etal~\cite{eyolfson2014correlations} studied the correlation between a commit's bugginess and its time-based properties (\eg commit's frequency, time of the day, day of the week) in three open-source datasets where they labeled \bugin commits by \bszz. Bernardi \etal~\cite{bernardi2018relation} used \agszz to label four open-source datasets so that they could analyze how a code change's fault proneness relates to the communication between developers who commit the code change. \etal~\citet{Izquierdo-Cortazar2012} and \citet{tufano2017empirical} investigated the correlation between developers' experience and defect proneness in the datasets they labeled by \bszz.

\subsection{The Multi-Commit Development Model} \label{prszz_sec:back_pr}
Multi-commit development models are code delivery systems that allow a developer to bundle changes organized across multiple commits. This practice is used to wrap features or fixes that benefit from being separated in different milestones.


This feature had a particular resonance in GitHub, with the Pull-Based development model. In this model, contributors clone a software project's repository to their local repository by using the \texttt{fork} function and make their changes in the source code independently of each other. When the code change is ready to be submitted, a contributor creates a pull request and the following iterative process begins: A core team member of the project (integrator) reviews the changes. If the changes are satisfactory, the integrator merges the pull request (\ie pulls the contributors' changes to the main branch); if the changes are unsatisfactory, the integrator may request additional changes or reject the pull request.

There are various options to merge pull requests.
By using the standard \texttt{merge} option, integrators merge a pull request into the main branch by retaining all the commits from the contributor's local branch (\ie preserving the history).
Similar to \texttt{merge}, \texttt{rebase} option integrates changes from one branch to another. However, \texttt{rebase} moves a feature branch (\eg contributor's local branch) to another branch (\eg master) without preserving the feature branch.
The \texttt{squash} option combines all the commits in a pull request into a single commit and merges this single commit to the main branch. As a result, a series of commits in a contributor's local branch corresponds to a single commit in the software project's master branch.

The choice of using these merging options depends on the policies and/or the model of multi-commit development used by a project. For example, in the approach used at \mozilla, the whole commit-set completing a development task is merged in a \texttt{rebase} fashion and, sometimes, a commit-set may be merged only partially (\eg merge commit A and B first, then C after a while).

GitHub is the most popular provider of hosting services for git projects. However, depending on the underlying Distributed Version Control System (DVCS), a project can be stored in a different remote location and supported with different tools to integrate commitments. For example, Mozilla has its custom source-code management tool that allows developers to choose the DVCS they prefer and apply all changes to the remote repository, which is a Mercurial repository: git users commonly create branches, while mercurial users rely on mercurial bookmarks.

\subsection{The Multi-Commit Model in Empirical SE}\label{prszz_sec:back_pr_relevance}
In 2014, the study by \citet{kalliamvakou2014promises} revealed that the majority of projects (65\%) used the multi-commit model, (\ie pull requests, PRs).
In 2016, the study by \citet{gousios2016work} emphasized an increasing trend in the usage of PRs, reporting that  over 135,000 repositories on GitHub received more than 600,000 PRs, and over 45\% of collaborative GitHub projects were pull-based.

The raw data for software quality studies often comes from the repositories (\eg GitHub) that support the multi-commit development model~\cite{Toth:2016, pascarella2019,Kim:2008,Wen:2016,yan2020}. For instance, Kim \etal~\cite{Kim:2008} and Wen \etal~\cite{Wen:2016} labeled the datasets that contain PRs by using \bszz to learn defect prediction models.
However, in an empirical study conducted on 398 releases of 38 Apache projects, \citet{herbold2019issues} showed that the existence of commit-sets results in the introduction of false positives on the labeled datasets (\ie wrongly identifying commits as \bugout commits). The reason for these false positives was that the \szz linking marked commits as \bugout in the wrong pull-request.

\subsection{Detecting Bugs at the Commit-Set Level}
Detecting defective commit-sets is a crucial activity for different stakeholders. In this respect, \citet{adams2016modern} highlights how Release Managers would benefit from tools able to detect defective commit-sets and prevent them from being integrated into upcoming releases. The presence of undiscovered bugs in a release may generate a cascade effect that contaminates branches created on top of the defective release or that import defective modifications. Besides the slowdown of code production in such branches, to locate a bug in a CI environment, each commit of the release needs to be tested, further increasing the workload on the bug resolution \cite{zampetti2019study}. In the work by \citet{castelluccio2019empirical}, Mozilla Release Managers state that based on the content, multiple commits (or patches) can be uplifted together and, for this reason, they are wrapped under the same commit-set (or release). Uplifting a commit-set means integrating all the commits in the production environment. Since Release Managers mostly rely on developers' suggestions to review a commit-set, skipping the stabilization phase increases the probability to introduce a bug, causing a regression in the codebase. 

Developers are interested in defect detection at coarser granularity too. For instance, the scope of the defect may not be limited to the modified lines as assumed by the defect predictors: ensuring quality standards in such cases requires an update of different connected files, as reported by \citet{dunsmore2003practical}. Also, analyzing a commit-set may give more insights into the root location of a specific bug \cite{ram2018makes}. Even more recently, Wan \etal~\cite{wan2018perceptions} interviewed practitioners to understand their expectations about the future evolution of defect prediction. On the topic of granularity, both experienced and beginner developers agree that defect prediction should extend the scope to commit-sets: despite the indication of bug location may be less precise, targeting commit-sets gives the reviewers a better overview of the code quality. In this way, the bug-fixing procedure increases the \textit{evolvability} of code components.

\section{Creating The \rosab}\label{prszz_sec:rosa_gt}

In 2021, \citet{pascarella2021evaluating} presented a novel approach to generate benchmarks to evaluate \szz at the commit level. Their approach relies on the fact that developers sometimes explicitly document---in the commit message---which commit introduced the defect they fix with the current commit (\eg ``THRIFT-4513: fix bug in comparator introduced by e58f75d''~\cite{pascarella2021evaluating}).
Starting from this bases, the methodology by \citet{pascarella2021evaluating} applies information retrieval techniques to filter commits' messages looking for an unequivocal link between a \bugin commit and a \bugout one.

Based on this novel approach and a further manual validation they conducted, \citet{pascarella2021evaluating} created and released a novel dataset gathering data from publicly available repositories on GitHub. Their dataset comprises 1,930 links between \bugin and \bugout commits pertaining to eight popular programming languages (C, C++, C\#, Java, JavaScript, Ruby, PHP, and Python). 

\subsection{Adapting the original dataset by \citet{pascarella2021evaluating} for evaluating \szz in the multi-commit development model}\label{prszz_sec:rosa_gt:adapt}

Being validated and approved by the software engineering research community for evaluating \szz in the single-commit context, we consider the dataset by \citet{pascarella2021evaluating} as a valuable ground to evaluate \szz in the multi-commit context as well.
Therefore, we adapted the dataset by \citet{pascarella2021evaluating} for this context. This adaptation is possible because the original dataset gathers data from GitHub projects, therefore commits can belong to pull requests (\ie commit-sets). Locating to which pull request a \bugout commit belongs as well as to which pull request the linked \bugin commit belongs makes it possible to define a \bugout commit-set and a \bugin commit-set, respectively.

 \begin{figure}[h]
	\includegraphics[width=\linewidth]{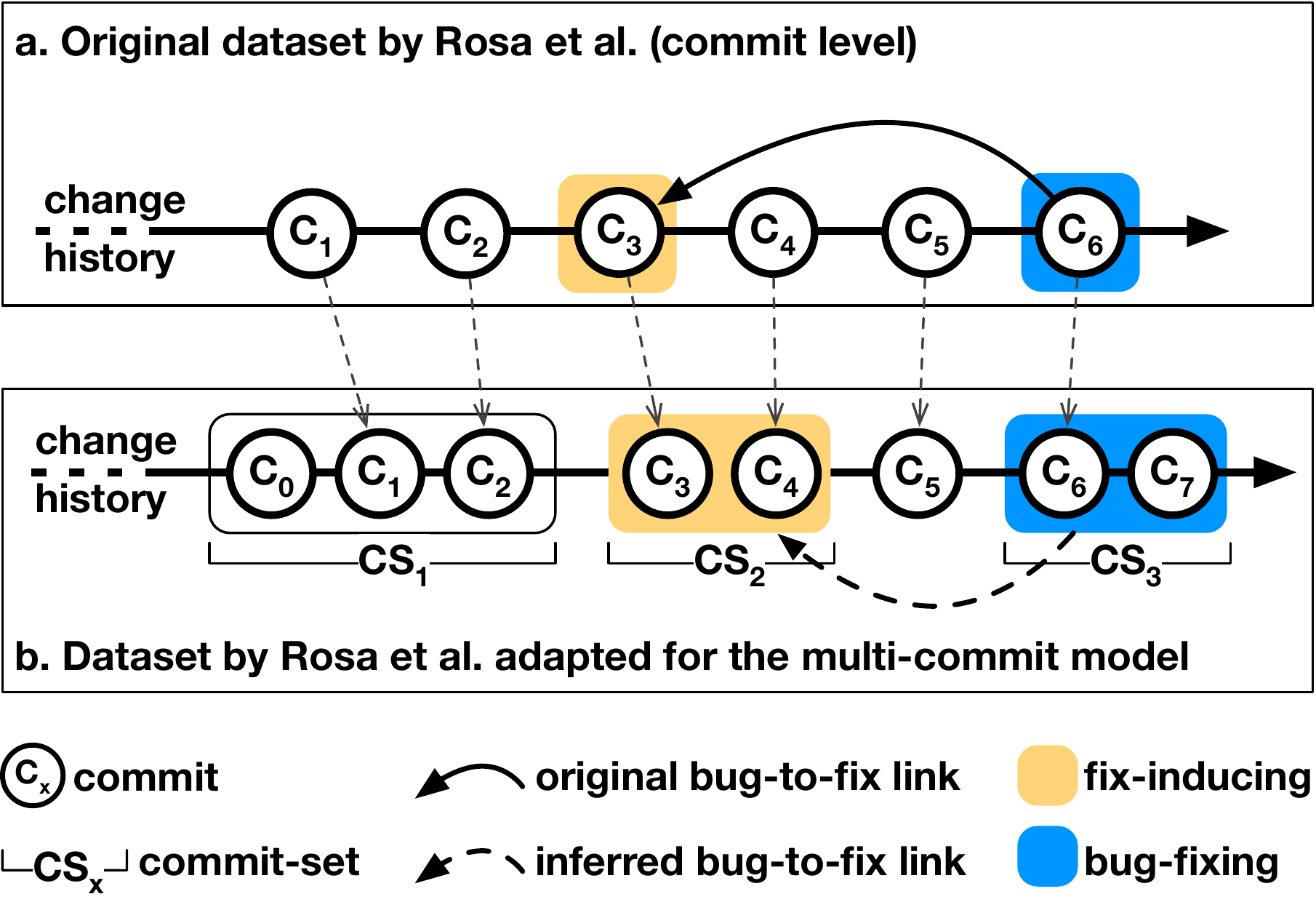}
	
	\caption{Adaptation of the \rosab from (a) commit level to (b) commit-set level (\ie in the multi-commit development model)}
	\label{fig:commit-to-cs}	
\end{figure}

\Cref{fig:commit-to-cs} shows this adaptation process. The top-half of the figure (\ie Figure 2a) shows the history of a software system with an example bug-to-fix link, as it is available in the original dataset by \citet{pascarella2021evaluating}. In particular, $c_6$ is a \bugout commit and is linked to $c_3$--its \bugin commit. Both commits can be linked to pull requests ($c_6$ to $CS_3$ and $c_3$ to $CS_2$). Then, $CS_3$ is labeled as a \bugout commit-set and $CS_2$ is its corresponding \bugin commit-set.

In practice, to implement this adaptation, we build a script that leverages GraphQL API \cite{GraphQL} to query commits' information from GitHub for each bug-to-fix link in the original dataset by \citet{pascarella2021evaluating}. Specifically, for each link in the dataset we:
\begin{itemize}
  \item extract the \bugin commit;
  \item use GraphQL API to check whether this \bugin commit belongs to a commit-set (\eg this happens for $c_3$ in \Cref{fig:commit-to-cs});
  \item in case a commit-set is found (\eg $CS_2$ in \Cref{fig:commit-to-cs}), add all the commits in the commit-set (\ie $c_3$ and $c_4$) to the list of \bugin commits for that bug-to-fix link.
  \item repeat the same for the \bugout commit.
\end{itemize}

When adapting the original dataset by \citet{pascarella2021evaluating} for our goal, we encountered the following problems:

\begin{description}[leftmargin=0.3cm]
  \item[Missing repositories.] Some commits in the dataset belong to repositories that are no longer available in GitHub (\ie they became private or were deleted).
 
  \item[Missing pull requests.] Most of the links include commits that cannot be linked to a PR. This case is also depicted in \Cref{fig:commit-to-cs} in the case of $c_5$: If $c_5$ was either a \bugin or \bugout commit, the entire link could not be considered because of the missing PRs. We had to discard these cases.
 
  \item[Forks.] GitHub produces incorrect mapping whenever the bug and fix are merged from a fork. A fork is a copy of a repository mostly used to perform custom changes to a project. In GitHub, changes from a fork can be imported in the main project through PRs. If the PR is applied in a fork and then imported in the original one, the PR \texttt{id} will change creating a duplicated reference to the same PR. This happens because the PR \texttt{id} is a progressive counter that in a fork is reset to 0. During the merge, GitHub assigns new PR \texttt{id}s to integrated PRs to keep the consistency with the main project \texttt{id} counter. Therefore, querying the PR \texttt{id} of a specific commit may return multiple references when a commit exists in a PR created in a fork and also integrated in the main project. This may lead to a confusing mapping with unrelated PRs or commits. Moreover, a given PR \texttt{id} can correspond to both a PR created in a fork and a different PR integrated in the original project. As a result, while inspecting a fork, GraphQL will return all commits (including unrelated commits) that belong to all PRs with the same reference \texttt{id}, if this PR is also in the original project.
\end{description}

Due to the aforementioned issues, we had to discard 100 links due to missing repositories, 1,680 links because the \bugin or the \bugout commit was not part of a pull request, and five links due to forks from the original dataset by \citet{pascarella2021evaluating}. The resulting adapted benchmark contains 145 links bug-to-fix commit-sets. Henceforth, we refer to this adapted dataset as \textbf{\rosab}. The 145 links have all 1:1 cardinality (\ie one \bugin commit-set and one \bugout commit-set) and comprise a total of 2,142 commits, of which 1,315 in \bugin commit-sets and 827 in \bugout ones.

\subsection{Limitations of \rosab}\label{prszz_sec:rosa_gt_limitations}

Despite the high quality bug-to-fix links provided by \citet{pascarella2021evaluating}, \rosab has the following limitations:

\smallskip
\noindent\textbf{Benchmark Size.} The adapted dataset, due to the issues encountered in the processing phase (\Cref{prszz_sec:rosa_gt:adapt}), only consists of 145 bug-to-fix links. This size poses limitations to the generalizability of the results.

\smallskip
\noindent\textbf{Ghost Commits and Extrinsic Bugs.} Different studies \cite{rezk2021ghost, rodriguez2020bugs} highlight the presence of bugs that cannot be retrieved by \szz. \citet{rezk2021ghost} refers to \textit{ghost commits} whenever a \bugin or a \bugout commit cannot be retrieved due to their modification type. In fact, a fix containing only new lines cannot be back-traced by Version Control System (VCS) log functions (\eg \texttt{git blame}) and a bug produced only by line deletions does not leave traces in future commits. Also, some modifications may be operated outside the VCS domain (\eg platform updates) yet produce a bug that requires a fix in the source code. For this reason, \szz produces false positive results on any processed commit that fixes such bugs. These cases are defined by \citet{rodriguez2020bugs} as \textit{extrinsic bugs}.

Both ghost commits and extrinsic bugs have a relevant impact in the real-case scenarios: \citet{rezk2021ghost} found out that, on average, the 15.78\% of all modifications are ghost commits, and \citet{rodriguez2020bugs} found that 15\% of bugs are extrinsic.

When analyzing the original dataset by \citet{pascarella2021evaluating}, we unexpectedly found none of these cases (even at the commit level) represented. That is, no linked commits were either a \textit{ghost commit} or an \textit{extrinsic bugs}.
This may limit the representativeness of \szz evaluations using this dataset.

\smallskip
\noindent\textbf{Candidate \bugin commits outside commit-sets.}
%
Even in the adapted dataset, \szz could output candidate commits that do not belong to any commit-set. \Cref{fig:szz_workflow} (b) shows such an example. \szz could mark commit $c_7$ as \bugin candidate, but $c_7$ is not part of any commit-set.
To make the evaluation of \szz results possible in these cases, we consider any candidate \bugin commit not belonging to a commit-set as belonging to a virtual commit-set composed of only the candidate commit itself.

\section{Creating the \mozillab}\label{prszz_sec:gt}

Many datasets have been proposed to test \szz, and even more have been built with \szz to train and study defect prediction models. However, none of them have been directly built by the code owners.
Also, \szz has never been studied at coarser granularity: despite a \bugout commit gathered by researchers belongs to a commit-set, only this specific commit is taken into account to retrieve the bug. This could be related to multiple factors: NLP techniques to detect keywords in \bugout commits may exclude other fix-related commits, or researchers' understanding of the content of commits may be limited due to the lack of experience with the project and its technologies.

In this regard, to properly investigate \szz and address limitations of \rosab presented in \Cref{prszz_sec:rosa_gt_limitations}, it is essential to have a reliable benchmark, which is possibly large-scale, based on real-world data, and relies on developers' work rather than researchers' approximations~\cite{pascarella2021evaluating}.

In this section, we describe how we created such a benchmark at \mozilla. Henceforth, we refer to this dataset as the \textbf{\mozillab}.



\subsection{Data Collection and Preparation}\label{prszz_sec:wf}
\begin{figure}[h]
	\centering
	\includegraphics[width=8cm]{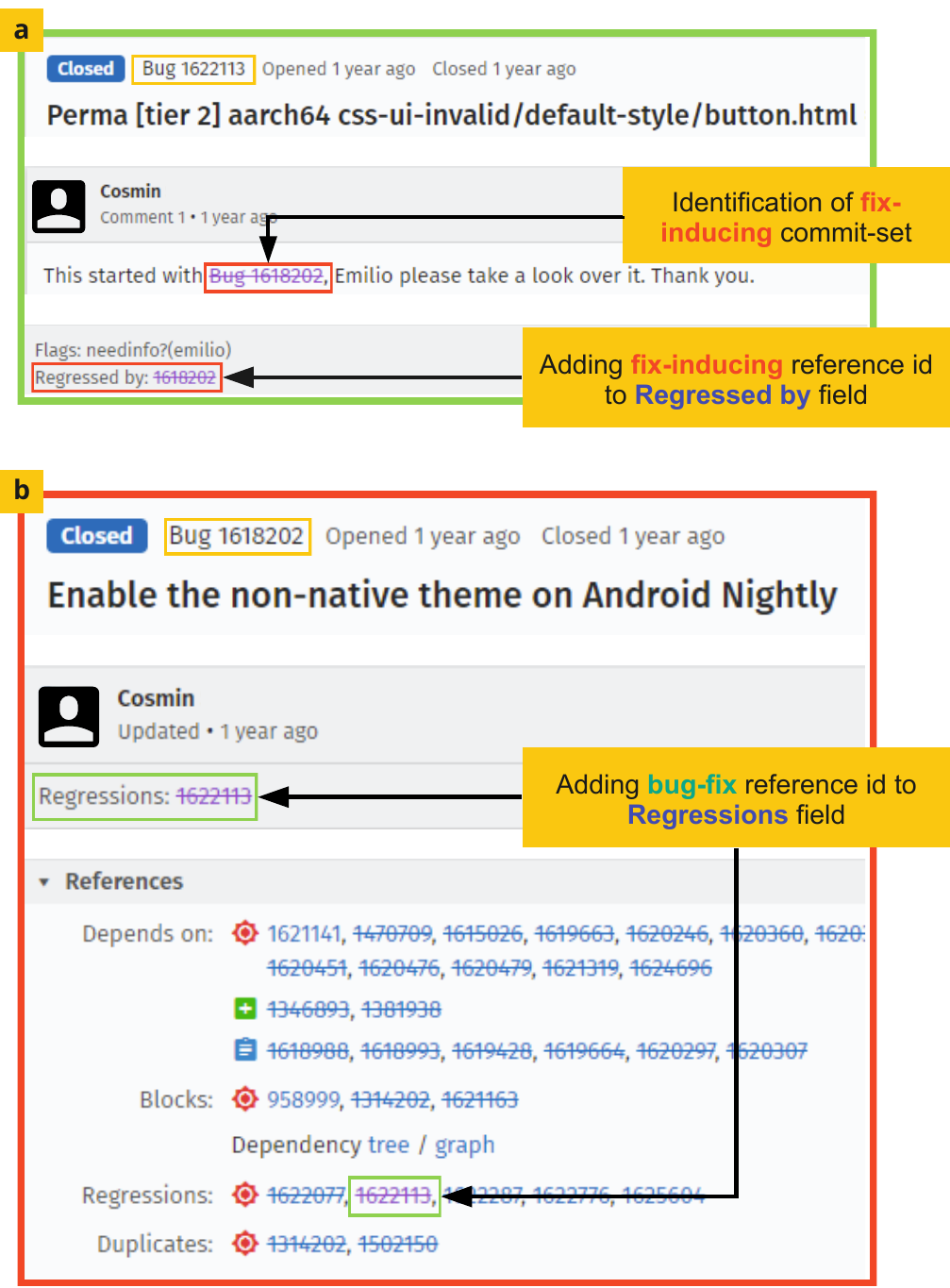}
	\caption{An example showing how developers fill in `\texttt{regression}' and `\texttt{regressed by}' fields in \issueTracker:  \BugzillaReports  for \textbf{(a)} \texttt{bug 1622113} and \textbf{(b)} \texttt{bug 1618202}.}
    \label{fig:regression-regressedby}
\end{figure}

The process that led to data in our benchmark consists of the following steps:

\begin{enumerate}[leftmargin=15pt]
  \item \textbf{Modification of \TheIssueTracker:} In April 2019, we created a dedicated field `\texttt{regressions}' and its mirror `\texttt{regressed by}' for each issue entry in  \TheIssueTracker. For each bug $b$, in the field `\texttt{regressed by},' there is the list of issues whose fixing commits introduced the bug $b$, while the field `\texttt{regressions}' contains the list of bugs that the fixing commits of bug $b$ itself introduced. For instance, as shown in \Cref{fig:regression-regressedby}, once developers detect that fixing \texttt{bug 1618202} introduced \texttt{bug 1622113}, they specified `\texttt{bug 1618202}' in the `\texttt{regressed by}'  field of \texttt{bug 1622113}. Also, `\texttt{bug 1622113}' is specified in the `\texttt{regression}'  field of \texttt{bug 1618202}.\smallskip

  \item \textbf{Data completion and collection in  \TheIssueTracker:} Since the introduction of the new fields in April 2019, filling in the `\texttt{regressions}'  and `\texttt{regressed by}' fields in issue reports has become a norm for \mozilla developers, users, and QA engineers whenever relevant and possible. The fields are validated by the developers who are assigned to either the regressing or the regressor issue. In addition, developers also filled in these fields for a subset of older issues dating backward to 2007.
  At the date of the creation of the benchmark used in our study, a total of \numOfTotalPRs issues were linked by \numDevelopers practitioners and verified by the developers assigned to the issues.\smallskip

  \item \textbf{Linking the commit-sets:}
  Each issue is fixed through a commit-set. Once the links among the \bugin and \bugout issues were established (through the previous step), we mapped the commit-sets solving these issues to one another to create the final dataset. To do so, we extended a tool at \Mozilla, BugBug \cite{bugbug}, to gather the relevant data from the issue tracker and code repository.
  
  
  
  We created a script to combine the information contained in the \BugzillaReports (\textit{regressed} and \textit{regressed by} flags) with commit information retrieved through the bug \texttt{id}. This way, we created the set of linked \bugin and \bugout commit-sets that also includes commit-related data.

\end{enumerate}

\subsection{\mozillab's Descriptives}\label{prszz_sec:gt_dataset}
Applying the procedure described in the previous subsection, we obtained a dataset of \numOfLinkedPRs links between \bugin and \bugout commit-sets, comprising \numOfLinkedCommits different commits and \numOfTotalPRs commit-sets. The relation between \bugout and \bugin commit-sets is $1:N$: For each bug there can be only one \bugout commit-set, but the bug could be induced in several commit-sets. This situation occurs in 121 cases. The cases where a bug is addressed by multiple \bugout commit-sets are extremely rare (only 9 cases), so we discarded them to avoid any incorrect mapping. A total of 1,586 pairs have a $1:1$ commit cardinality ratio. In the dataset, 7,043 commits are \bugout, 16,159 commits are \bugin, and 884 commits are in a chain of both \bugout and \bugin commit-sets. Even though we collected a snapshot to generate the dataset used in the current study, the `\texttt{regressions}' and `\texttt{regressed by}' fields we devised and deployed are still in use at \Mozilla and the dataset is continuously growing with new data. To the best of our knowledge, the dataset we contribute with this paper is currently the largest publicly available dataset with \bugout and \bugin links among commits and commit-sets.

Our benchmark is based on commit-sets whose code belongs to the \MozillaLong's codebase. This codebase represents a heterogeneous system employing a variety of programming languages and application contexts, ranging from web development to statistical analysis.
The heterogeneity of \Mozilla's codebase contributes to (1) increasing the variety in the nature of the cases on which we apply \szz and (2) reducing the bias introduced by focusing on a specific programming language or domain. 

\begin{table}[ht]
	\begin{center}
		
		\caption{Languages involved in the \Mozilla codebase}
		\label{tab:project_languages}
		{\rowcolors{2}{LightCyan}{white}
		\begin{tabular}{l|r|r|r|r}
			\textbf{Language} & \textbf{Files} & \textbf{Blanks} & \textbf{Comments} & \textbf{LOC}\\\midrule
			\textbf{JavaScript} & 72,870 & 1,199,781 & 1,753,236 & 5,540,827 \\
			\textbf{C++} & 11,772 & 801,098 & 669,043 & 4,476,606 \\ 
			\textbf{HTML} & 90,776 & 463,590 & 105,185 & 4,118,159 \\ 
			\textbf{C/C++ Header} & 16,564 & 519,357 & 956,346 & 2,475,718 \\ 
			\textbf{Rust} & 8,365 & 246,505 & 442,208 & 2,384,387 \\ 
			\textbf{C} & 3,998 & 321,980 & 502,674 & 2,158,164 \\ 
			\textbf{JSON} & 2,245 & 883 & 0 & 1,190,423 \\ 
			\textbf{Python} & 6,746 & 222,750 & 260,302 & 872,281 \\ 
			\textbf{XML} & 2,813 & 7,005 & 2,973 & 453,026 \\ 
			\textbf{Assembly} & 561 & 35,477 & 35,924 & 294,756 \\ 
			\textbf{INI} & 12,582 & 73,130 & 175 & 231,725 \\ 
			\textbf{XHTML} & 3,562 & 23,033 & 8,097 & 189,678 \\ 
			\textbf{Java} & 854 & 24,503 & 62,588 & 156,493 \\
			\textbf{Other} & 11,893 & 745,785 & 178,693 & 1,227,722 \\\midrule
			\textbf{Total} & 251,601 & 4,174,520 & 4,977,444 & 25,769,965 		\end{tabular}
	    }
	\end{center}		
\end{table}

\section{Research Questions}\label{prszz_sec:rqs}


We set to empirically evaluate how \szz performs when applied to commit-sets. 
Therefore, we ask:

\begin{center}
	\begin{rqbox}
		\begin{description}
			\item[RQ1:] What is the performance achieved by \szz and its main variations at the commit-set level? 
		\end{description}
	\end{rqbox}
\end{center}

The change of granularity leads to a change in the input space of \szz: Each \bugout commit-set can be composed of more than one commit, as opposed to at the commit level where only one input commit can be used. This may have both positive and negative effects: Increasing the algorithm's input could increase the chances to find \bugin candidates, yet it could also lead to more false positives. 

We hypothesize that by removing the irrelevant and noisy commits from a \bugout commit-set, the overall results of \szz would significantly improve.
In our second research question, we set out to investigate this hypothesis and, if confirmed, study whether and to what extent an automated approach can automatically recognize non-useful commits from a \bugout commit-set. Finally, we use this approach to re-compute the new results for an improved \szz at commit-set level. Therefore, we ask:

\begin{center}
	\begin{rqbox}
		\begin{description}
			\item[RQ2:] To what extent can \szz's results be improved by retaining only the useful input commits from a \bugout commit-set?
		\end{description}
	\end{rqbox}
\end{center}

\section{RQ1: \szz performance for commit-sets} \label{prszz_sec:performances}

In this section, we evaluate the performance of \szz when applied to the multi-commit context.

\subsection{Methodology}


We evaluate the performance of \szz at commit-set level from three complementary perspectives.

\begin{description}[leftmargin=0.3cm]

  \item[- Evaluation perspective 1:] We consider the \mozillab. Given a \bugout commit-set, we run \szz on each commit it includes. Then, for each \bugin candidate commit found, we consider as output all the commit-sets that include at least one of these candidate commits.
  For example, in \Cref{fig:szz_workflow}, we consider both $CS_1$ and $CS_2$ as the output, because they each contain at least one commit that was linked by \szz from the commits in $CS_3$.

\end{description}


\begin{description}[leftmargin=0.3cm]

  \item[- Evaluation perspective 2:] We consider the cases in \mozillab in which both the fix and the bug consists of a single commit. This selection corresponds to recasting the evaluation to match a commit level one and enables a more direct comparison with, for example, the work by \citet{pascarella2021evaluating}. \mozillab contains 1,586 such cases.

  \item[- Evaluation perspective 3:] We consider \rosab. In this subset of the original dataset by \citet{pascarella2021evaluating} that we created (see \Cref{prszz_sec:rosa_gt}), both the \bugin and \bugout commits are embedded in commit-sets. This enables a meaningful comparison at the commit-set level.
  \rosab contains 145 such cases.

\end{description}

\begin{table*}[]
\begin{center}
		
		\caption{Characteristics of the datasets considered in the different perspectives}
		\label{tab:descriptives_of_perspectives}
		{\rowcolors{3}{LightCyan}{white}
\begin{tabular}{l|p{1.5cm}|p{4cm}|r|r|r|r|r|r}
\multirow{2}{*}{\textbf{Perspective}} & \multirow{2}{*}{\textbf{Dataset}} & \multirow{2}{*}{\textbf{Characteristics}}                                                                                                                                                                                                   & \multicolumn{3}{c|}{\textbf{Number of links}} & \multicolumn{3}{c}{\textbf{Number of commits}}                         \\
                             &    &                                                                                                                                                                                                                                & \textbf{total}      & \textbf{1:1}      & \textbf{1:N}      & \textbf{total} & \textbf{in \bugin set} & \textbf{in \bugout set} \\\midrule
\textbf{Perspective 1}                & \mozillab & Created by Mozilla developers (\Cref{prszz_sec:gt})                                                                                                                                                                                                   & 5,348 & 5,227 & 121 & 24,086 & 7,927 & 17,043 \\
\textbf{Perspective 2}                & Subset of \mozillab & A single commit on both \bugin and \bugout commit sets                                                                                                       & 1,586 & 1,586 & 0 & 2,869 & 1,586 & 1,358 \\
\textbf{Perspective 3}                & \rosab & Subset of the dataset by \citet{pascarella2019} where both \bugin and \bugout commits are in a commit set & 145 & 145 & 0 & 2,142 & 827 & 1,315                                         
\end{tabular}}
\end{center}
\end{table*}

\Cref{tab:descriptives_of_perspectives} provides information on how the different perspectives are related to the datasets we consider.\smallskip

\noindent\textbf{Evaluation metrics.}
To evaluate the performance of \szz, we adopt measures of recall, precision, and F1 score as used in information retrieval~\cite{manning2010introduction}:

\begin{equation*}
recall = \frac{|correct \cap identified|}{|correct|}
\end{equation*}

\begin{equation*}
precision = \frac{|correct \cap identified|}{|identified|}
\end{equation*}

\begin{equation*}
F1 = 2 \cdot \frac{recall \cdot precision}{recall + precision}
\end{equation*}

\smallskip

In the formulas above, \textit{identified} represents the set of candidate commit-sets retrieved by \szz, and \textit{correct} represents the set of \bugin commit-sets established by developers in the \mozillab or by \citet{pascarella2021evaluating}.

Furthermore, we compute the Jaccard distance. This measure represents the similarity between two sets as the proportion of shared elements among all elements in both sets. Given a specific commit-set ($CS$), we consider as sets for the Jaccard distance (1) the \bugin commits retrieved by \szz ($FiC_{\szz}$) and (2) the \bugin commits from the ground truth ($FiC_{gt}$).

\begin{equation*}
JD(CS_i) = 1 - \frac{|FiC_{\szz}(CS_i) \cap FiC_{gt}(CS_i)|}{|FiC_{\szz}(CS_i) \cup FiC_{gt}(CS_i)|}
\end{equation*}

\begin{equation*}
JD(variation) = \frac{1}{n}\displaystyle\sum_{i=1}^{n}JD(CS_i)
\end{equation*}

A Jaccard distance's value closer to 1 means a higher dispersion between the two considered sets and closer to 0 is almost no dispersion.\smallskip

\noindent\textbf{\szz variations.} We test several variations of \szz focusing on non-language-specific ones: \bszz, \agszz, \rszz, and \lszz. We use the implementation of these algorithms as offered by \citet{pascarella2021evaluating}. Moreover, we include in our evaluation the improved version of \bszz provided by \pydriller~\cite{spadini2018pydriller}, which---differently from the original \bszz---allows users to specify a set of commits to exclude. 
The developers from \mozilla created a list of massive-refactoring commits (\eg \Mozilla commit 7558c\footnote{\protect\url{https://hg.mozilla.org/mozilla-central/rev/7558c8821a074b6f7c1e7d9314976e6b66176e5c}}), which we filtered out using \texttt{git-hyper-blame} function when applying \szz. This makes PyDriller's variation of \bszz similar to \raszz. \smallskip

\subsection{Results}
In the following, we present the results of our evaluations by perspective.\smallskip

\begin{table}[h]
    \caption{\szz's commit-set level performance}
    \label{tab:temps}
    \begin{subtable}[h]{0.45\textwidth}
        \centering
        \caption{Perspective~1 (\mozillab), $N = 5,348$}
		\label{tab:szz_performances}
		{\rowcolors{2}{LightCyan}{white}
		\begin{tabular}{l|r|r|c|c|c|c}
			\textbf{Variation} & \textbf{Identified} & \textbf{Correct} & \textbf{Rec.} & \textbf{Prec.} & \textbf{F1} & \textbf{avg JD}
			\\\midrule
			\textbf{B-SZZ} & 13,323 & 2,582 & 0.49 & 0.19 & 0.28 & 0.81\\
			\textbf{AG-SZZ} & 7,964 & 1,997 & 0.38 & 0.25 & 0.30 & 0.82\\ 
			\textbf{L-SZZ} & 4,975 & 1,527 & 0.29 & 0.31 & 0.30 & 0.84\\ 
			\textbf{R-SZZ} & 5,008 & 1,981 & 0.38 & 0.40 & 0.39 & 0.80\\ 
			\textbf{PyDr.} & 13,420 & 2,548 & 0.48 & 0.19 & 0.27 & 0.81\\ 
		\end{tabular}
		}
    \end{subtable}
    \newline
    \vspace*{0.5 cm}
    \newline
    \begin{subtable}[h]{0.45\textwidth}
        \centering
        \caption{Perspective~2 (subset of the \mozillab with a single commit on both \bugin and \bugout commit-sets), $N = 1,586$}
		\label{tab:szz_commit_performances}
		{\rowcolors{2}{LightCyan}{white}
		\begin{tabular}{l|r|r|c|c|c|c}
			\textbf{Variation} & \textbf{Identified} & \textbf{Correct} & \textbf{Rec.} & \textbf{Prec.} & \textbf{F1} & \textbf{avg JD}
			\\\midrule
			\textbf{B-SZZ} & 1,310 & 763 & 0.47 & 0.58 & 0.52 & 0.65\\
			\textbf{AG-SZZ} & 1,236 & 620 & 0.38 & 0.50 & 0.43 & 0.66\\ 
			\textbf{L-SZZ} & 1,245 & 470 & 0.29 & 0.37 & 0.33 & 0.70\\ 
			\textbf{R-SZZ} & 1,239 & 626 & 0.39 & 0.50 & 0.44 & 0.61\\ 
			\textbf{PyDr.} & 1,291 & 754 & 0.47 & 0.58 & 0.52 & 0.65\\
		\end{tabular}
		}
     \end{subtable}
     \newline
    \vspace*{0.5 cm}
    \newline
    \begin{subtable}[h]{0.45\textwidth}
        \centering
        \caption{Perspective~3 (\rosab), $N = 145$}
		\label{tab:szz_rosa_pr_performances}
		{\rowcolors{2}{LightCyan}{white}
		\begin{tabular}{l|r|r|c|c|c|c}
			\textbf{Variation} & \textbf{Identified} & \textbf{Correct} & \textbf{Rec.} & \textbf{Prec.} & \textbf{F1} & \textbf{avg JD}
			\\\midrule
			\textbf{B-SZZ} & 861 & 76 & 0.51 & 0.09 & 0.15 & 0.36\\
			\textbf{AG-SZZ} & 499 & 71 & 0.47 & 0.12 & 0.22 & 0.38\\ 
			\textbf{L-SZZ} & 256 & 52 & 0.36 & 0.21 & 0.26 & 0.41\\ 
			\textbf{R-SZZ} & 247 & 68 & 0.47 & 0.35 & 0.35 & 0.40\\ 
			\textbf{PyDr.} & 676 & 55 & 0.37 & 0.13 & 0.13 & 0.37\\ 
		\end{tabular}
		}
     \end{subtable}

\end{table}

\begin{description}[leftmargin=0.3cm]

  \item[Evaluation perspective 1]-
  \Cref{tab:szz_performances} shows the results of this perspective.
  \rszz is the variation reaching the highest F1 measure, matching the ranking reported by \citet{pascarella2021evaluating} at the commit level.
  In terms of recall, \bszz and \pydriller outperform the other variations. Most likely because there is no filtering of the considered commits, which, as a trade-off, lowers the precision significantly.
  
  The last column of \Cref{tab:szz_performances} shows the results in terms of average Jaccard distance between the ground truth and \szz variations. The Jaccard distance is always higher than 0.8, highlighting a remarkable dispersion of the results.

  \item[Evaluation perspective 2]-
  This perspective's results (\Cref{tab:szz_commit_performances}) show an increase in precision with negligible decreases in recall.
  For commit-sets of one commit only, \bszz and \pydriller represent the best choice.
  
  In terms of average Jaccard distance (last column of \Cref{tab:szz_commit_performances}), we observe that the average dispersion of results is high but generally lower than in commit-set case. This triangulates the improvements in terms of precision providing further evidence that having only one input commit for \szz significantly reduces the dispersion from the expected result set, when compared to having more than one input commit.

  \item[Evaluation perspective 3]-
  \Cref{tab:szz_rosa_pr_performances} reports the results for perspective 3.
  Similarly to the case of perspective 1, \rszz outperforms every other model, and \bszz as well as \pydriller have low precision due to more input commits.
  Yet, the performance of \szz, when applied to this subset of \rosab, is better compared to when \szz is applied to the \mozillab.
  This result is confirmed by the average Jaccard distance, which is the lowest among the three perspectives.

\end{description}

Overall, in all three perspectives, the performance of \szz (for all the variations we tested) at commit-set level is 20 or more percentage points lower in both precision and recall than that reported by \citet{pascarella2021evaluating} at the commit level.
For example, in our dataset \bszz reaches 0.49 and 0.19 for recall and precision, respectively, while it reaches 0.69 and 0.39 in the study by \citet{pascarella2021evaluating}.

Since we use a different dataset than \citet{pascarella2021evaluating}, we cannot rule out that the reason for this lower performance is due to specific characteristics of the \mozillab, rather than the change in granularity from commit to commit-set level. Indeed, the results for Perspective 2 (which simulates a commit level scenario, thus makes it more comparable to the study by \citet{pascarella2021evaluating}) in which even the best performing variation does not achieve the results reported by \citet{pascarella2021evaluating}, may indicate this as a likely reason. At the same time, the results presented in Perspective 3 (\ie evaluated on \rosab) are also lower: It could be that commit-set level poses specific challenges hindering the effectiveness of \szz.

\begin{center}
	\begin{rqbox}
		\begin{description}[leftmargin=0cm]
			\item[Finding 1:]
			The performance of \szz and its variations is 20 percentage points lower than the results previously reported at the commit level.
		\end{description}
	\end{rqbox}
\end{center}

\subsection{Further Analysis of the Results}
To understand whether and how specific conditions influence \szz, we further analyzed cases in which variations behaved similarly or very differently.
In particular, we analyzed the cases of commit-sets
\begin{enumerate*}[label=(\Alph*)]
  \item missed by all variations (\ie false negatives),\label{enum:all_miss} and
  \item missed or retrieved by only one \szz variation.\label{enum:one_only}
\end{enumerate*}
Our aim is to investigate whether these commit-sets share particular features that can be used in the future to improve \szz selection criteria. In this respect, we focused on results obtained by \szz on \mozillab as we consider this ground truth more reliable because it is entirely built by Mozilla's developers. \smallskip

\begin{figure*}
     \centering
     \begin{subfigure}[b]{0.33\textwidth}
     \centering
     \includegraphics[width=\textwidth]{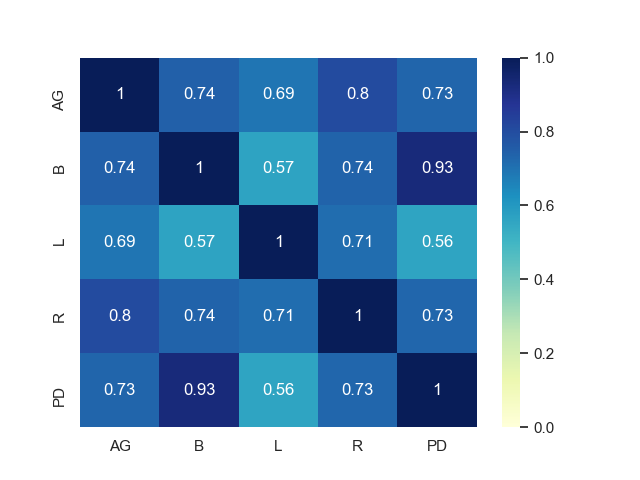}
         \caption{True Positives}
         \label{fig:positive_overlap}
     \end{subfigure}
     \hfill
     \begin{subfigure}[b]{0.33\textwidth}
         \centering
         \includegraphics[width=\textwidth]{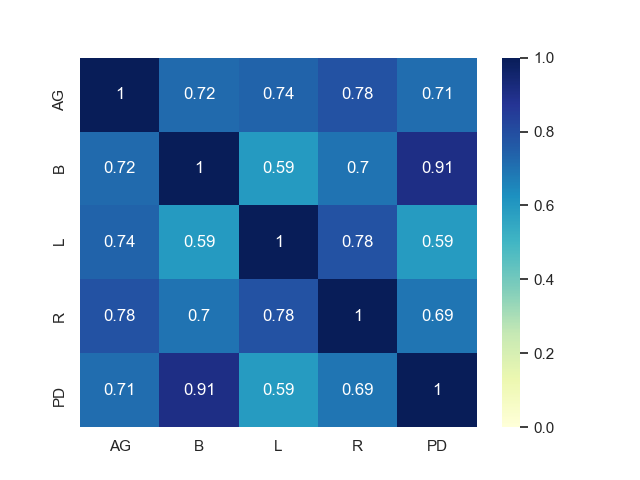}
         \caption{False Positives}
         \label{fig:negative_overlap}
     \end{subfigure}
     \hfill
     \begin{subfigure}[b]{0.33\textwidth}
         \centering
         \includegraphics[width=\textwidth]{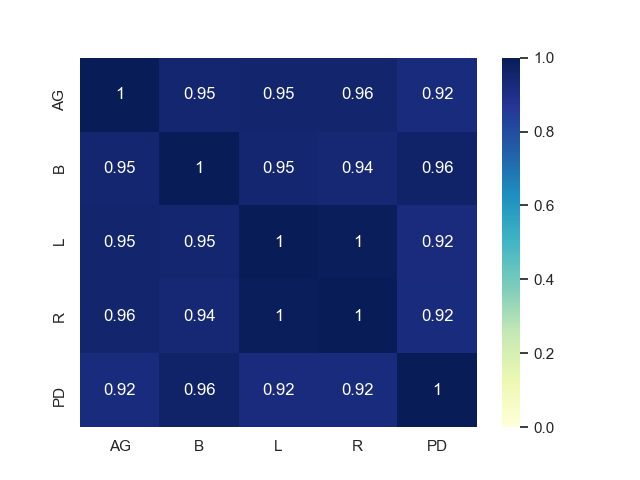}
         \caption{False Negatives}
         \label{fig:undiscovered_overlap}
     \end{subfigure}
        \caption{Results overlapping among variations by \szz implementation pairs}
        \label{fig:overlap}
\end{figure*}

\noindent\textbf{\ref{enum:all_miss} Commit-sets missed by all variations.}
\Cref{fig:overlap} shows an overview of the agreement among pairs of \szz variations, in terms of true positives (\ie the agreement rate on correct results - \Cref{fig:positive_overlap}), false positives (\ie the agreement rate on incorrectly identified commit-sets - \Cref{fig:negative_overlap}), and false negatives (\ie the agreement rate on missed commit-sets - \Cref{fig:undiscovered_overlap}).
In \Cref{fig:positive_overlap} and \Cref{fig:negative_overlap}, the agreement is never lower than 56\%. This is expected, since all the algorithms heavily derive from the \bszz root.
In \Cref{fig:undiscovered_overlap} (the agreement rate for missed commit-sets), the agreement is as high as 90\%.

In total, we have 1,174 links in our dataset that cannot be retrieved by \szz. Those links represent the 45\% of the entirely mismatched cases and the 22\% of the \dataset.

One of the authors manually inspected randomly selected 50 cases of false negatives. For each of them, he examined the nature of the changes in both the \bugin and \bugout commit-sets (which lines and files are modified, what kind of modifications have been performed) and debugged \szz execution to detect the root cause of the error. In this way, we have been able to spot the following three main reasons:

\begin{itemize}[label={--}]

  \item \emph{New lines of code cause mismatches} - 
  Whenever a bug is fixed by only introducing new lines of code, the algorithm cannot retrieve the \bugin commit. We found 21 of these cases in our \dataset. Moreover, 10 of them also do not share any file with the \bugin commit-set. This limitation is known at commit-level as the \textit{ghost commit effect}~\cite{da2016framework, rezk2021ghost}.

  \item \emph{A bug was introduced in a different file} - 
  The main reason why \szz cannot reach a \bugin commit is that the \bugout commit-set modifies different files from the \bugin commit-set. This situation is present in \mozillab in 1,164 cases (43\% of the entirely mismatched cases). In 367 of them, the files in the \bugout commit-set and the ones in the \bugin commit-set do not even share the same directory. Such cases still belong to the definition of \textit{Ghost Commits}~\cite{da2016framework, rezk2021ghost}.

  \item \emph{No Commit in \bugin commit-set} -
  In 45 cases we found that there is no \bugin commit-set associated to the bug. This condition usually happens when some updates to the environment are applied but such modifications are not operated by code (\eg a library update). In this case, \szz cannot be applied since it produces only false positive results. However, the algorithm cannot prevent such occurrences. This kind of bugs are known in literature as \textit{Extrinsic Bugs}~\cite{rodriguez2020bugs}. As an example, the bug 1525373\footnote{\url{https://bugzilla.mozilla.org/show_bug.cgi?id=1525373}} has its root in the previous fix of the bug 1569091\footnote{\url{https://bugzilla.mozilla.org/show_bug.cgi?id=1569091}}. In this case, the problem was created by a local machine with an inner problem not related with Mozilla's code. To solve that, developers needed to reinstall the Operating System (OS) on that machine, generating a compatibility problem with the Mozilla's codebase. To solve the bug, the developers added a few instructions to enable the new OS to instrument the code. So, for this reason, there is a \bugout commit-set (to address the compatibility issue) but the \bugin problem (the new installed OS) is not reflected on the VCS.

\end{itemize}

\smallskip
\noindent\textbf{The Case of the \rosab.}
To understand how prominent the aforementioned three conditions are in other cases, we checked the percentage of untraceable links (\ie \textit{Ghost Commits} and \textit{Extrinsic Bugs}) in the original dataset by \citet{pascarella2021evaluating} at both commit and commit-set level. Unexpectedly, \emph{none of the ground truth links in the dataset (and, as a consequence, in \rosab) is affected by any of the three aforementioned conditions}: the bug is always placed in at least a commit and the fix always includes at least a modification/deletion in a file that is shared by the \bugin and \bugout commits.

This characteristic of the dataset by \citet{pascarella2021evaluating} (which is reflected also on \rosab) is surprising because past research (as well as the data in the \mozillab) has provided evidence that the three conditions exist in the majority of situation.
The lack of these cases in the dataset by \citet{pascarella2021evaluating} may be due to reasons connected to the methodology with which the benchmark was created, yet there is no self-evident reason we could discover.

\begin{table}[h]
	\begin{center}
		
		\caption{\szz's commit-set level performance on \mozillab when excluding non-linkable cases ($N = 4,174$)}
		\label{tab:szz_performances_filtered}
		{\rowcolors{2}{LightCyan}{white}
		\begin{tabular}{|l|r|r|r|r|r|r|}
			\textbf{Variation} & \textbf{Identified} & \textbf{Correct} & \textbf{Rec.} & \textbf{Prec.} & \textbf{F1} & \textbf{avg JD}
			\\\midrule
			\textbf{B-SZZ} & 10,730 & 2,573 & 0.63 & 0.24 & 0.35 & 0.54\\
			\textbf{AG-SZZ} & 6,342 & 1,986 & 0.48 & 0.31 &  0.38 & 0.56\\ 
			\textbf{L-SZZ} & 4,028 & 1,515 & 0.37 & 0.38 & 0.37 & 0.62\\ 
			\textbf{R-SZZ} & 4,050 & 1,970 & 0.48 & 0.49 & 0.48 & 0.49\\ 
			\textbf{Pydriller} & 10,814 & 2,515 & 0.61 & 0.23 & 0.34 & 0.54\\ 
		\end{tabular}
		}
	\end{center}		
\end{table}

To quantify the effect of excluding \textit{Ghost Commits} and \textit{Extrinsic Bugs}, we discarded cases that cannot be linked in our \dataset and re-run the experiments. \Cref{tab:szz_performances_filtered} reports the results.

As expected, we observe a strong increase of \szz performance in each of its variations. In most of the cases, \szz also performs better in \mozilla \dataset than in the \rosab. After removing these unlinkable cases, the ranking of the variations is stable: \rszz outperform every other model in terms of precision and F1 score, but \bszz and \pydriller reach higher recall.

\begin{center}
	\begin{rqbox}
		\begin{description}[leftmargin=0cm]
			\item[Finding 2:]
			More than 20\% of the links in \mozillab cannot be retrieved by current implementations of \szz. By excluding these cases, the recall values of \szz at commit-set level improve substantially, getting closer to those reported for the commit level.
		\end{description}
	\end{rqbox}
\end{center}

\smallskip
\noindent\textbf{\ref{enum:one_only} Commit-sets retrieved/missed by only one variation.}
To gain further insights into the conditions under which \szz works/fails, we focus on the commit-sets that were correctly identified (115 cases) or missed (547) by \emph{only one} \szz variation.

For each \szz variation, we randomly extracted a statistically significant sample (for a total of 262 records) and manually analyzed each case, exploring the git history and defining why a given variation behavior differs from the others.
Two authors performed this analysis independently: they only agreed on which sources to consider (bug reports and source code) during the analysis. Then, each of the authors described the problem case in a few words, specifying the reason behind the unique miss or finding related to the \szz variation and its heuristic.
Their initial results reached an inter-rater agreement of 82.82\%.
Then, the two authors discussed the cases of disagreement until they reached an agreement.

In the following, we report the findings by variation.\smallskip

\begin{description}[leftmargin=0.3cm]

  \item[\bszz:] The true positive findings unique to \bszz are often identified through the lines with code comments. Comment lines help \bszz succeed where other variations fail because the latter discard such lines \emph{a-priori}. Also, some bugs reside in large refactoring commits, but only the basic version of \szz can detect this type of bug.
  
  At the same time, refactoring commits represent the primary source of error for \bszz~\cite{neto2018impact}.

  \item[\agszz:] This variation relies on an annotation graph to connect modified lines with functions or methods that wrap them, providing a better mapping of the code and excluding cosmetic changes. However, in our experimental setup, it does not provide any particular benefit in identifying \bugin commits: we are not able to detect special cases where \agszz outperforms any other variation. Also, we found that \agszz incorrectly labeled the correct \bugin changes as refactoring. This incorrect labeling leads the algorithm to retrieve as \bugin changes commits older than the correct ones (in particular, when multiple files or commits are involved). Furthermore, the criteria to detect refactoring commits are sometimes too stringent, inducing \agszz to mark specific commits as irrelevant.
  
  A typical example of \agszz failure is represented by the bug \texttt{1668755}\footnote{\url{https://bugzilla.mozilla.org/show_bug.cgi?id=1668755}}: here the \bugout commit-set is composed by only one commit, \texttt{a7aef33}\footnote{\url{https://github.com/ambroff/gecko/commit/a7aef3341915becacfc6a0edc961461b147da896}}, and the \bugin commit-set is represented by the commit \texttt{4413dfb}\footnote{\url{https://github.com/ambroff/gecko/commit/4413dfbe1a9b9d4d78104cfb8d1aedfeb087c875}}. The \bugin and \bugout commits share the file \texttt{defaultBrowserNotification.js} but, since the modification contains over 150 modified lines, it is considered a refactoring. For this reason, \agszz traced the file to an earlier point, wrongly landing on the commit \texttt{d261b6a}\footnote{\url{https://github.com/ambroff/gecko/commit/d261b6a4f26055b719dc23302ca4033ce3ed3f8d}} (\ie the first commit where the file appeared).

  \item[\lszz \& \rszz:] These variations filter the set of \bugin candidates only keeping the largest (\lszz) and the most recent (\rszz) commits. We found that only in seven cases, the two variations perform better than the other variations in detecting \bugin commits.
  On the contrary, they have the highest amount of unique errors (\eg \lszz has 377 unique errors). These mistakes are caused by the selection criteria of \lszz and \rszz. In the case of \rszz, the algorithm often stopped before reaching the correct commit marking more recent commits as bug-inducing. For instance, this often happened when the correct \bugin commit was followed by a refactoring commit. 
  \lszz, instead, marks as \bugin the commit that contains the highest number of changes. However, this criterion often led to mistakes: \eg \lszz selected a refactoring commit (where a large number of lines of code have been changed) as opposed to the correct \bugin commit (where only few lines have been modified).

  \item[\pydriller:] This variation can be considered as an extended version of \bszz since it provides the possibility to specify a list of commits to skip during the execution. In fact, most of the \bugin commits correctly identified only by \pydriller are located in changes that other \szz variations marked as refactoring.
  For instance, \pydriller outperformed the other variations when identifying errors concerning modifications in the parameters of a method or related to code that undergoes multiple cosmetic changes. 
  In its execution, \pydriller skips all the commits marked as refactoring by \Mozilla developers (see \Cref{prszz_sec:performances}). However, Mozilla's developers listed only a portion of commits that could be considered as refactoring. For this reason, \pydriller is not able to skip some big changes (unrelated to the bug), thus providing incorrect results. 
  Moreover, the implementation of \pydriller (version 1.15) we considered cannot correctly identify C/C++ directives (\eg \texttt{\#ifndef} or \texttt{\#ifdef}) marking them as comments.

\end{description}

\begin{center}
	\begin{rqbox}
		\begin{description}[leftmargin=0cm]
			\item[Finding 3:]
			Filtering \szz output increases the precision of the algorithm, sometimes discarding useful commits. Such filtering heuristics can be leveraged in respect of the trade-off between precision and recall.
		\end{description}
	\end{rqbox}
\end{center}

\section{RQ2: Selection of Input Commits}

When applied at the commit-set level, \szz and its variations achieve low precision values (\eg see \Cref{tab:szz_performances}).
The underlying reason might be that some commits in a \bugout commit-set are irrelevant to the bug fix and create additional noise for \szz.
This hypothesis is corroborated by the results of evaluation Perspective 2 (\ie the analysis of 1-commit commit-sets, shown in \Cref{tab:szz_commit_performances}): In the cases in which only one commit is available as input for \szz, the precision of the algorithm is significantly higher, with only minor losses in terms of recall.

In this research question, we first set out to challenge our hypothesis by evaluating whether and to what extent removing all irrelevant/noisy commits from the input of \szz improves its performance at commit-set level.

Based on results supporting our hypothesis, we investigate to what extent an automated approach can classify commits in a \bugout commit-set as either useful or not for \szz linking. Last, we evaluate the final effects of using such an approach on the results of \szz.

\subsection{The Impact of Irrelevant Commits on \szz}
To test our hypothesis, we use the \rosab considering the cases in which both \bugin and \bugout commits are within commit-sets (\ie cases used for evaluation Perspective 3 in RQ1). The use of this benchmark helped us to get an overview from many different projects. 

We remove all the commits in each \bugout commit-set that lead to \szz linking wrong \bugin commit-sets. Then, we evaluate the performance of \szz and compare to those achieved without this filtering (\ie \Cref{tab:szz_rosa_pr_performances}) to verify our hypothesis.

\subsubsection{Data Labeling}
Whether a commit in a \bugout commit-set is a good linker depends on the \szz variation.
For example, a commit could lead to a wrong commit when using \bszz, yet to the correct \bugin commit when used as input for \pydriller.
Therefore, we apply two labeling methods and evaluate their effect:

\begin{itemize}
	\item \textbf{\singlevarLabeling}: We label commits based on a specific \szz variation. For example, when evaluating \bszz, we consider as good all the commits that provide good links for \bszz, while we consider as bad the commits that do not help \bszz (even if they help other variations).
	\item \textbf{\allvarLabeling}: We consider as good links only \bugout commits that lead to correct links for \emph{all} \szz variations, and we consider as bad links only \bugout commits that provide a bad link for \emph{all} \szz variations. We exclude commits that lead to good/bad links for only a subset of variations.
\end{itemize}

These two labeling options also influence the dataset we consider: in the \singlevarLabeling, we can consider all commits in the computation, while in the \allvarLabeling, we rely only on overlapping results, thus reducing the available amount of \bugout commits.

\subsubsection{Results}

\begin{table*}[t]
	\begin{center}
		
		\caption{\szz performance upper-bound at the commit-set level considering perspective~3 for \singlevarLabeling, $N = 145$}
		 \label{tab:nobad_commits}
		{\rowcolors{2}{LightCyan}{white}
		\begin{tabular}{|l|r|r|c||r|r|r|}
			\hline
	   \textbf{Algorithm} & \textbf{Identified} & \textbf{Correct} & \textbf{Discarded Commit-sets} & \textbf{Prec.} & \textbf{Rec.} & \textbf{F1} \\
			\hline\hline
			\textbf{B-SZZ} & 259 & 76 & 70/145 & 0.29 & 0.51 & 0.37 \\
			\textbf{AG-SZZ} & 144 & 71 & 74/145 & 0.49 & 0.48 & 0.48 \\ 
			\textbf{L-SZZ} & 52 & 52 & 93/145 & 1.00 & 0.36 & 0.53 \\ 
			\textbf{R-SZZ} & 68 & 68 & 77/145 & 1.00 & 0.47 & 0.64 \\ 
			\textbf{Pydriller} & 159 & 55 & 91/145 & 0.34 & 0.37 & 0.35 \\\hline
   			\textbf{Overlap} & 35 & 35 & 110/145 & 1.00 & 0.24 & 0.39 \\
   
			\hline
		\end{tabular}
		}
	\end{center}		
\end{table*}

\Cref{tab:nobad_commits} reports the results of \szz considering only commits in \bugout commit-sets that provide a correct link to a \bugin commit-set adopting the evaluation perspective 3. Those results represent the highest performance achievable by each \szz variation, maximizing the precision of the algorithm. We consider them as an upper bound to our machine learning pipeline.
\rszz is the best performing model, followed by \lszz. The key difference between these variations and the rest is the precision: they both reach 1.00, meaning they perfectly identify the \bugin commit-set without any false positives. For the same reason, \bszz completely lacks in precision: the broader range of \bugin candidates implies a higher number of correct links (and thus a higher recall) and incorrect \bugin candidates.
By filtering the input commits of \szz, we will increase the precision of algorithms like \bszz and \pydriller, reducing the amount of identified commit-sets and trying to keep the recall as high as possible. Also, there is no way to improve the recall of such algorithms without modifying their internal behavior. Reducing the number of \bugout commits to consider in a commit-set may also reduce the number of correct links retrieved, thus reducing the recall of the algorithm. We aim to find the right trade-off to increase the precision more than the amount of lost recall.

\begin{center}
	\begin{rqbox}
		\begin{description}[leftmargin=0cm]
			\item[Finding 4:]
			Filtering \szz input commits increases the algorithm's precision, but it is impossible to improve the recall without designing a new \szz variation. 
		\end{description}
	\end{rqbox}
\end{center}

\subsection{Automated Classification of Input Commits for \szz}
The results in \Cref{tab:nobad_commits} support our hypothesis: \szz performs better when the input is reduced to only the relevant commits. Moreover, the additional improvement is significant, thus justifying the investigation on how to select and keep only the relevant commits in a commit-set. Based on this finding, we present our investigation on the creation of an automated approach to classify commits into good/bad for \szz linking. 

\subsubsection{The Methodology in a Nutshell}
The automated approach needs to tackle a binary classification problem: Given a commit in a \bugout commit-set, it determines whether it is a \emph{good linker} commit that can be used by \szz to find the correct \bugin commit-set.

Employing sophisticated techniques (\eg based on deep learning) to accomplish this task goes beyond the scope of the current work. Our aim is two-fold: (1) Verifying whether it is
feasible to create an automatic classification approach that provides reasonable results and (2) defining an initial baseline against which future methods can be tested.
Therefore, we employ supervised machine learning, which bases its decision on a set of features we define for which the weights and interaction are automatically computed from the training set.

We take advantage of the \mozillab to train and test several machine learning algorithms with the features we define. Then, we run a \emph{cross-dataset evaluation}: We train our models on the \mozillab and evaluate their results on the data from \rosab. This cross-dataset evaluation allows us to better establish the generalizability of the results, because we apply the classifier to an entirely unseen dataset, extracted from different projects with diverse development practices.

\subsubsection{Dataset Creation}
As the purpose of the machine learning approach is to classify \bugout \textit{commits} into good or bad links for \szz, we need to create a dataset at the commit level granularity. To this aim, we gathered all commits from all \bugout commit-sets in the \mozillab. We got 7,048 data points from the 5,348 commit-sets in the \mozillab.

\begin{table*}[t]
	\begin{center}
		
		\caption{Feature extracted in the \bugout commits of the dataset}
		\label{tab:szz_ml_features}
		{\rowcolors{2}{LightCyan}{white}
		\begin{tabular}{|l|l|p{3.5cm}|p{6cm}|l|}
			\hline
			\textbf{Feature Name} & \textbf{Granularity} & \textbf{Description} & \textbf{Rationale} & \textbf{type}
			\\\hline\hline
			\textbf{Addition} & Commit & Number of added LOC & Lines added in a commit are parts of the input of \szz. The more lines we have, the wider the search space and thus the chances to retrieve a \bugin commit. Also, \szz variations that rely on a heuristic to spot refactoring may find it beneficial to have a certain range of added lines. & Int \\
			\textbf{Deletion} & Commit & Number of deleted LOC & Lines deleted in a commit are parts of the input of \szz. Deleted lines can also be backtraced by \szz, and so the more deleted lines we have, the wider the research space. & Int \\ 
			\textbf{Files} & Commit & Number of modified files & The more files are modified, the more files \szz will explore. This feature represents the branching of \szz across a project. & Int \\ 
			\textbf{CS Addition} & Commit-set & Percentage of added LOC relative to the commit-set & As for added lines but in proportion with the entire commit-set. A higher percentage of added lines may indicate that most of the changes are condensed in a single commit that could highlight a good linker. & Float \\ 
			\textbf{CS Deletion} & Commit-set &Percentage of deleted LOC relative to the commit-set & As for deleted lines, but in proportion with the entire commit-set. If a commit contains most of the deletion, it means the code has been discarded for some reason, possibly a bug too. That may indicate that at a huge percentage of deleted lines corresponds to a good linker. & Float \\ 
			\textbf{CS Files} & Commit-set & Percentage of touched files relative to the commit-set & As for files, but in proportion with the entire commit-set. If a commit contains multiple modifications on different files, it may indicate the presence of a unlocalized bug. Therefore, we believe that the more files are involved, the more the commit is prone to contain a bug. Also, the commit with the highest percentage of modified lines has a broader branching scope that the others commit. & Float \\ 
			\textbf{Order} & Commit-set & Normalized time position in the commit-set & Like for \rszz, the order of commits may identify a relevant fix. By this heuristic, we can assume that the core fix is usually performed first and so the first commit of a commit-set is most probably the closest to the \bugin commit-set. & Float \\ 
			\textbf{CS Shared Files} & Commit-set & If any file modified is touched somewhere else & Extends the concept of bug locality to commit-set level~\cite{rahman2011bugcache}. If a file is modified multiple times in the same commit-set, it may indicate a certain impact/relevance of such file in the \bugout commit-set. & Bool \\ 	
			\hline
		\end{tabular}
		}
		
	\end{center}		
\end{table*}

\smallskip
\noindent\textbf{Classification Features.} Since we set to classify commits (as either useful for linking or not), we have the opportunity to compute features from both the given commit and the commit-set that includes the commit.
\Cref{tab:szz_ml_features} lists all the features we considered, their granularity, and our rationale for their inclusion.

\smallskip
\noindent\textbf{Cross-dataset Evaluation.} To perform the aforementioned cross-dataset evaluation, we also collected 827 data points from the \rosab.

\subsubsection{Training and Evaluation}

We train and evaluate the following classification models:
Support Vector Machine (SVM),
Decision Trees (DT),
Random Forest (RF),
Logistic Regression (LR),
Gradient Boost (GB), and
Naive Bayes (NB).
We selected these models because they are the most widely-used supervised algorithms in software engineering \cite{zhang2003machine} and make different assumptions about the underlying data and the interactions among the features.

Since our dataset is imbalanced, we combined those models with different sampling methods: Random Oversampling, Random Undersampling, and SMOTE \cite{chawla2002smote}.
To ensure the validity of our final results, we applied these sampling methods only to the training set.
Also, we used cost-sensitive learning \cite{ling2008cost} whenever possible, applying proportional weights based on labels' class ratio.
To obtain a comprehensive and more reliable set of results, we combined all models with different cross-validation techniques: shuffling, k-fold, and repeated k-fold validation with different numbers of splits/iterations. We scaled the features by unit variance to favor the convergence of SVM classifiers. Then, we evaluated precision, recall, F1 score, confusion matrix, and AUC-ROC for each model.

We also studied feature correlation with Spearman's Index to spot (and eventually discard) highly correlated features.
We also investigated feature importance to determine the features with the strongest impact on these models' prediction performance.
\Cref{tab:szz_ml_performances} reports the best combination of classifier, sampling method, and validation technique for each \szz variation. 

To corroborate our findings, we applied these models using cross-dataset evaluation: we used trained models to predict good linkers from \rosab \bugout commits in the version containing commit-set information.

\subsubsection{Results}
\textbf{Dataset Composition.}
The dataset in which we perform \allvarLabeling consists of 6,178 \bugout commits: 4,789 are labeled as bad-links while 1,389 are labeled as good-links. Instead, the \singlevarLabeling dataset contains 7,048 \bugout commits but the labeling ratio changes based on the \szz variation we are considering.

We found no strong correlation between features, thus no further refinement is necessary. We also conducted Principal Component Analysis (PCA) \cite{abdi2010principal} that revealed a contribution of almost all features to the variance: we cover the 95\% of dataset variance with five features (\feat{Deletion}, \feat{Files}, \feat{CS Addition}, \feat{CS Deletion}, and \feat{CS Shared Files}).

\textbf{Machine Learning Models.}
We tested several machine models that better address the problem of binary classification. Considering that the dataset is imbalanced, we rely on the F1 score to select the best models for each \szz variation: the false positive rate for highly imbalanced datasets decreased due to a large number of true negatives, and this undermines the reliability of both AUC and accuracy scores \cite{saito2015precision}.

\Cref{tab:szz_ml_performances} shows that all selected models perform similarly. In all variation-aware models, the optimum one is Gradient Boost which reaches 0.66 of F1 score and 0.69 in accuracy on \bszz. Also, in \pydriller we obtain relatively good performance, confirming the existing similarity between those two \szz variations. In the overlap approach, the best model is Logistic Regression. Although the F1 score in the overlap approach is in line with variation-aware models, we can detect higher accuracy and AUC scores. \Cref{tab:szz_ml_feature_importance} lists all contributions obtained by each feature in each model. The percentage of deleted lines is a good predictor for good linkers. 

\begin{center}
	\begin{rqbox}
		\begin{description}[leftmargin=0cm]
			\item[Finding 5:]
			\szz Machine Learning Models are a viable option to filter commit-set to detect the best bug-fix linker commits. The most relevant feature is the percentage of deleted lines in such commits with respect to the total deleted lines in the commit-set.
		\end{description}
	\end{rqbox}
\end{center}

\subsection{Evaluating \szz with Automatically Filtered Commits} \label{sec:rq2:szz_evaluation}

As a last step in our investigation, we use the best performing models (as trained and tested on \mozillab) to predict good bug linkers in the unseen \rosab. Then we filtered \szz results to keep only algorithm iterations on such commits, discarding all iterations performed on bad bug linkers. We applied both \singlevarBased and \allvarBased models, as reported in \Cref{tab:szz_rosa_pr_performances_ml_sv} and \Cref{tab:szz_rosa_pr_performances_ml_av} respectively. As expected, the \szz performances on the ground-truth at the commit-set level increase in precision while the recall is slightly reduced. However, if we add the Machine Learning filter on top of \szz first stage (\Cref{prszz_sec:back_szz_how}), the output obtained by the algorithm will be more reliable in terms of correctness. In fact, by excluding bad linker commits from the ground truth, the links between \bugin and \bugout commit-sets contain fewer false positives, and the relation between relevant and retrieved commit-sets is preferable. \rszz is still the most reliable solution for \bugin commit-set retrieval, with a minor increase of F1 score from 0.39 in the normal case to 0.41 for the \singlevarBased solution and 0.40 in the \allvarBased solution. \rszz performances are even better if we exclude bad-linkers also from the evaluation (with F1 score = 0.58 for \allvarBased solution) but, despite the higher scores, most of the issues have been completely excluded by \szz execution. This is because all \bugin commits of such commit-sets are labeled as bad linkers. In \singlevarBased models, the number of excluded commit-sets varies from 58 to 72, depending on the \szz variation, while we discard 77 commit-sets on the \allvarBased models. This is expected since the overlap solution is trained on stricter labeling constraints. However, although the machine learning models halve the number of the overall issue (145), the number of relevant commit-sets retrieved by \szz is quite similar. In the \singlevarBased configuration, we retrieve on average 19 relevant \bugin commit-sets less than the normal case. In comparison, on average, we miss 24.8 relevant \bugin commit-sets less in the \allvarBased configuration. This means that most discarded issues would produce incorrect results on \szz anyway.

\begin{center}
	\begin{rqbox}
		\begin{description}[leftmargin=0cm]
			\item[Finding 6:]
			Despite being imperfect, applying automated filtering of commits within commit-sets leads to an overall improvement in \szz at the commit-set level.
		\end{description}
	\end{rqbox}
\end{center}

\begin{table*}[t]
	\begin{center}
		
		\caption{Performance of the best machine learning models as computed on \mozillab}
		\label{tab:szz_ml_performances}
		{\rowcolors{2}{LightCyan}{white}
		\begin{tabular}{|l|c|l|l|r|c|c|c|c|c|c|}
			\hline
			\textbf{Algorithm} & \textbf{Model} & \textbf{Sampling} & \textbf{Cross Validation} & \textbf{Splits} & \textbf{Prec.} & \textbf{Rec.} & \textbf{F1} & \textbf{AUC} &\textbf{Acc.}
			\\\hline\hline
			\textbf{Overlap} & LR & SMOTE & N-Fold & 5 & 0.44 & 0.79 & 0.56 & 0.63 & 0.72 \\
			\textbf{\bszz} & GB & Random Oversampling & N-Time N-Fold & 5 & 0.54 & 0.82 & 0.66 & 0.57 & 0.69 \\
			\textbf{\agszz} & GB & Random Oversampling & Shuffling & 5 & 0.44 & 0.84 & 0.57 & 0.51 & 0.66 \\
			\textbf{\lszz} & GB & Random Oversampling & Shuffling & 5 & 0.35 & 0.81 & 0.49 & 0.47 & 0.64 \\
			\textbf{\rszz} & GB & Random Oversampling & N-Time N-fold & 5 & 0.43 & 0.81 & 0.56 & 0.54 & 0.64 \\
			\textbf{\pydriller} & GB & Random Undersampling & N-Fold & 10 & 0.53 & 0.83 & 0.65 & 0.68 & 0.67 \\
			\hline
		\end{tabular}
		}
	\end{center}		
\end{table*}
\begin{table*}[t]
	\begin{center}
		
		\caption{Feature importance for the best models as computed on the \mozillab.}
		\label{tab:szz_ml_feature_importance}
		{\rowcolors{2}{LightCyan}{white}
		\begin{tabular}{|l|c|c|c|c|c|c|c|c|}
			\hline
			\textbf{Algorithm} & \textbf{Addition} & \textbf{Deletion} & \textbf{Files} & \textbf{CS Addition} & \textbf{CS Deletion} & \textbf{CS Files} & \textbf{Order} & \textbf{CS Shared Files}
			\\\hline\hline
			\textbf{Overlap(*)} & 0.05 & 0.02 & -1.05 & -0.07 & \textbf{1.35} & 0.10 & 0.00 & 0.33 \\
			\textbf{\bszz} & 0.03 & 0.04 & 0.01 & 0.04 & \textbf{0.73} & 0.02 & 0.10 & 0.01 \\
			\textbf{\agszz} & 0.03 & 0.04 & 0.08 & 0.03 & \textbf{0.68} & 0.02 & 0.07 & 0.02 \\
			\textbf{\lszz} & 0.07 & 0.10 & 0.08 & 0.05 & \textbf{0.59} & 0.03 & 0.04 & 0.01 \\
			\textbf{\rszz} & 0.04 & 0.03 & 0.05 & 0.05 & \textbf{0.71} & 0.03 & 0.05 & 0.01 \\
			\textbf{\pydriller}  & 0.04 & 0.07 & 0.02 & 0.04 & \textbf{0.72} & 0.03 & 0.05 & 0.00 \\
			\hline
		\end{tabular}
		}
	\end{center}
	(*): since feature importance cannot be calculated for all models, we added the given contribution of weighted coefficients in Logistic Regression.
\end{table*}

\begin{table*}[t]
	\begin{center}
		
		\caption{Results of the cross-dataset evaluation with  \singlevarLabeling, $N = 827$}
  \label{tab:rosa_ml_sv_commit_selection}
		{\rowcolors{2}{LightCyan}{white}
				\begin{tabular}{|l|c|c|c|c|r|r|r|}
			\hline
           \textbf{Algorithm} & \textbf{TP} & \textbf{TN} & \textbf{FP} & \textbf{FN} & \textbf{Rec.} & \textbf{Prec.} & \textbf{F1} \\
			\hline\hline
			\textbf{\bszz} & 48 & 701 & 22 & 56 & 0.46 & 0.68 & 0.55 \\
   			\textbf{\agszz} & 48 & 715 & 24 & 40 & 0.54 & 0.66 & 0.60 \\
   			\textbf{\lszz} & 36 & 726 & 33 & 32 & 0.52 & 0.52 & 0.52 \\
   			\textbf{\rszz} & 43 & 708 & 27 & 49 & 0.46 & 0.61 & 0.53 \\
   			\textbf{\pydriller} & 47 & 703 & 21 & 56 & 0.45 & 0.69 & 0.54 \\
   
			\hline
		\end{tabular}
		}
	\end{center}		
\end{table*}

\begin{table*}[t]
	\begin{center}
		
		\caption{Results of the cross-dataset evaluation with \allvarLabeling, $N = 827$}
  \label{tab:rosa_ml_av_commit_selection}
		{\rowcolors{2}{LightCyan}{white}
				\begin{tabular}{|l|c|c|c|c|r|r|r|}
			\hline
           \textbf{Algorithm} & \textbf{TP} & \textbf{TN} & \textbf{FP} & \textbf{FN} & \textbf{Rec.} & \textbf{Prec.} & \textbf{F1} \\
			\hline\hline
			\textbf{\bszz} & 44 & 702 & 21 & 60 & 0.42 & 0.67 & 0.52 \\
   			\textbf{\agszz} & 43 & 717 & 22 & 45 & 0.48 & 0.66 & 0.56 \\
   			\textbf{\lszz} & 32 & 726 & 33 & 36 & 0.47 & 0.49 & 0.48 \\
   			\textbf{\rszz} & 41 & 711 & 24 & 51 & 0.44 & 0.63 & 0.52 \\
   			\textbf{\pydriller} & 45 & 704 & 20 & 58 & 0.43 & 0.69 & 0.53 \\
   
			\hline
		\end{tabular}
		}
	\end{center}		
\end{table*}

\begin{table*}[t]
	\begin{center}
		
		\caption{Performance of \szz at the commit-set level considering \rosab on perspective~3 (cases with both \bugin \& \bugout commits within commit-sets) for \singlevarLabeling, $N = 145$}
		\label{tab:szz_rosa_pr_performances_ml_sv}
		{\rowcolors{2}{LightCyan}{white}
		\begin{tabular}{|l|r|r|c||r|r|r||r|r|r|}
			\hline
			 & & & & \multicolumn{3}{c}{\textbf{Ground Truth Scores}} & \multicolumn{3}{c}{\textbf{Without Bad Linkers}} \\
	   \multirow{-2}{*}{\textbf{Algorithm}} & \multirow{-2}{*}{\textbf{Identified}} & \multirow{-2}{*}{\textbf{Correct}} & \multirow{-2}{*}{\textbf{Discarded Commit-sets}} & \textbf{Prec.} & \textbf{Rec.} & \textbf{F1} & \textbf{Prec.} & \textbf{Rec.} & \textbf{F1} \\
			\hline\hline
			\textbf{B-SZZ} & 113 & 53 & 58/145 & 0.44 & 0.34 & 0.38 & 0.44 & 0.54 & 0.49 \\
			\textbf{AG-SZZ} & 99 & 53 & 64/145 & 0.51 & 0.34 & 0.40 & 0.51 & 0.53 & 0.52 \\ 
			\textbf{L-SZZ} & 65 & 37 & 72/145 & 0.57 & 0.25 & 0.35  & 0.57 & 0.40 & 0.47 \\ 
			\textbf{R-SZZ} & 65 & 44 & 62/145 & 0.68 & 0.30 & 0.41 & 0.68 & 0.48 & 0.56 \\ 
			\textbf{Pydriller} & 104 & 52 & 64/145 & 0.47 & 0.33 & 0.39 & 0.47 & 0.54 & 0.51 \\ 
			\hline
		\end{tabular}
		}
	\end{center}		
\end{table*}

\begin{table*}[t]
	\begin{center}
		
		\caption{Performance of \szz at the commit-set level considering \rosab on perspective~3 (cases with both \bugin \& \bugout commits within commit-sets) for \allvarLabeling, $N = 145$}
  \label{tab:szz_rosa_pr_performances_ml_av}
		{\rowcolors{2}{LightCyan}{white}
				\begin{tabular}{|l|r|r|c||r|r|r||r|r|r|}
			\hline
            & & & & \multicolumn{3}{c}{\textbf{Ground Truth Scores}} & \multicolumn{3}{c}{\textbf{Without Bad Linkers}} \\
	   \multirow{-2}{*}{\textbf{Algorithm}} & \multirow{-2}{*}{\textbf{Identified}} & \multirow{-2}{*}{\textbf{Correct}} & \multirow{-2}{*}{\textbf{Discarded Commit-sets}} & \textbf{Prec.} & \textbf{Rec.} & \textbf{F1} & \textbf{Prec.} & \textbf{Rec.} & \textbf{F1} \\
			\hline\hline
			\textbf{B-SZZ} & 101 & 45 & 77/145 & 0.45 & 0.30 & 0.36 & 0.45 & 0.52 & 0.48 \\
			\textbf{AG-SZZ} & 86 & 44 & 77/145 & 0.51 & 0.30 & 0.38 & 0.51 & 0.51 & 0.51 \\ 
			\textbf{L-SZZ} & 60 & 33 & 77/145 & 0.55 & 0.22 & 0.32  & 0.57 & 0.40 & 0.47 \\ 
			\textbf{R-SZZ} & 60 & 42 & 77/145 & 0.70 & 0.28 & 0.40 & 0.70 & 0.49 & 0.58 \\ 
			\textbf{Pydriller} & 98 & 46 & 77/145 & 0.47 & 0.31 & 0.37 & 0.47 & 0.53 & 0.50 \\ 
			\hline
		\end{tabular}
		}
	\end{center}		
\end{table*}


\section{Discussion}\label{prszz_sec:discussion}

\noindent
\textbf{A Commit-Set \szz is needed.} The multi-commit development model, in general, and commit-sets, in particular, have become a widely adopted practice in software development~\cite{gousios2016work}. Our results show that \szz seems not to be ready for this more complex context. For this reason, further research should refine \szz for the multi-commit model.
Our study revealed that, when applied to commit-sets, \rszz outperforms \szz variations, similarly to the commit-level case. However, \bszz and \pydriller outclass all other variations in terms of recall. This means that the noise introduced by the coarser granularity may represent an obstacle, but at the same time, the extension of the search space represents a benefit in terms of the \bugin commits retrieved.
Reducing the noise by performing a commit selection in the \bugout commit-set is a viable option. As proposed in this work, a machine learning model can represent a powerful tool in this perspective, but further investigation is needed to discover new features that can increase ML model performances.

\smallskip
\noindent
\textbf{Integrating dynamic project information in \szz.} Our manual analysis confirmed existing problems of \szz: the impossibility for the algorithm to establish a relationship between \bugin and \bugout commits when they do not have files in common~\cite{rezk2021ghost, rodriguez2020bugs}.
To resolve such cases, we envision integrating \szz with information on the dynamic flow of the project, as only the code's static analysis currently performed by \szz is not sufficient in this context.
However, this kind of information might prove challenging to extract, especially for non open-source repositories. To solve this issue, we propose to integrate this information directly into \TheIssueTracker, where \szz could easily access it.

\smallskip
\smallskip
\textbf{Benchmark.} The results of our evaluation of \szz and its variations, considering all Commit-Sets and the Commit-Sets with only one commit, present significant differences from what was reported in the study by Rosa \etal~\cite{pascarella2021evaluating} (as shown in \Cref{prszz_sec:performances}). These differences might have been caused by the different datasets used for the evaluation. Project-specific aspects (\eg the programming languages used) or practices might introduce bias in the performance of \szz when applied to a specific dataset. 
For this reason, we believe future studies should focus on evaluating \szz (and its variations) using multiple datasets to mitigate possible biases. To this aim, we made our dataset openly available to be used in future evaluations of \szz.

\section{Threats to Validity}\label{prszz_sec:TTV}

\noindent
\textbf{Construct Validity.} As mentioned in section \ref{prszz_sec:gt}, we deployed our dedicated data field in April 2019. Based on the information retrieved from \TheIssueTracker, developers improved the description of most bugs from 2007 to date. However, the link between bug and fix is not always explicitly available: especially when bugs are trivial to fix, the bug discussion tends to be short or nonexistent. Although original developers cared about their bugs, it is still possible that recalling information about the specific bug could be challenging. This could have led to missing or false links in our benchmark.

The development process at \Mozilla matches the use of \texttt{rebase} strategy to merge a commit-set into the main branch. For this reason, all commits belonging to the same commit-set appear in the history of a single, unique branch. Actually, we are not able to gather any evidence on how \szz performs in combination with \texttt{merge} or \texttt{squash} operations. However, this condition favors the application of \szz, considering its limitation due to the use of annotation functions. Having a linear and unified environment is the only opportunity to keep the commit history consistent without losing the possibility to run \szz along the entire code evolution of \Mozilla's projects.

\smallskip
\noindent
\textbf{Internal Validity.}
Carrying out a manual analysis usually introduces a subjectivity in the output. Also in our case, we cannot assure the full correctness of our findings in Section \ref{prszz_sec:performances}. However, the fact that two authors reached the 82.82\% of inter-rater agreement before any confrontation highlights a certain evidence of conditions affecting \szz performances. Also, the rigorous process of inspection of the git-history, supported by the comparison with other variations' results, increases our confidence in the findings.

As mentioned in \Cref{prszz_sec:gt_dataset}, our dataset belongs to a complete new generation. Considering the different granularity, a comparison with previous dataset may result unfeasible. In this respect, we did our best to compare our work with Rosa's one by flattening the problem from both the granularities: the commit level and the commit-set level. Also, we voluntarily excluded ghost commits and extrinsic bugs from our dataset to fairly represent the effect of numerous filtering stages adopted by Rosa \etal~ \cite{pascarella2021evaluating}.

\smallskip
\noindent
\textbf{External Validity.} \Mozilla is a heterogeneous case of study under many aspects: it involves multiple technologies and programming languages, and is aimed at different contexts like security, web development, machine learning, and data analysis. However, \Mozilla is an open source project with high standards in development practices and code quality. For this reason, it is not representative of all development contexts.
To mitigate the effect this bias may have on the results, we also considered \rosab, which gathers data from a large set of open-source software systems hosted on GitHub.

\section{Conclusions}\label{prszz_sec:conclusion}

We evaluated the performance of \szz and its variations (\ie \szz algorithms) in a multi-commit environment. We extended the work by \citet{pascarella2021evaluating} to analyze the problem at the commit-set granularity. We also designed and deployed a new dedicated data field in the \mozilla \TheIssueTracker, which  \numOfLinkingDevs \Mozilla developers and QA engineers used to link \numOfLinkedPRs bug-fixing issues to \bugin issues across \linkingDuration months (and these numbers are still increasing).

As a result of conducting quantitative and qualitative analyses to evaluate \szz algorithms' performance, we found that \rszz achieves the best performance, whereas \szz variations proposed for commit-level also apply in a multi-commit environment. Moreover, machine learning models can effectively increase the precision of \szz at coarser granularity and help exclude \bugout commits that would lead \szz to an incorrect result.

Overall, the main contributions of this paper include:

\begin{itemize}[leftmargin=0.5cm]
  \item A publicly available dataset of \numOfLinkedPRs links between \bugin and \bugout commit-sets (totaling \numOfLinkedCommits commits), whose creation involved professional developers from \Mozilla;
  \item An empirical evaluation of \szz and its variations at commit-set level using both Rosa's and our dataset, showing that \rszz is the most reliable solution beyond the change of granularity;
  \item Empirical data, based on a manual investigation of 262 commit-sets, on the unique findings/mistakes of each considered \szz variation that highlights their shortcomings in this context;
  \item Empirical evidence of the impact of irrelevant commits in commit-sets on the results of \szz at the commit-set level.
  \item The creation and evaluation of a set of machine learning models to automatically detect commits that are good linkers for \szz in a commit-set, as well as an evaluation of their application on the final results of \szz.
\end{itemize}

	\appendices
	
	\bibliographystyle{IEEEtranN}
	\bibliography{bib}

\begin{thebibliography}{50}
\providecommand{\natexlab}[1]{#1}
\providecommand{\url}[1]{#1}
\csname url@samestyle\endcsname
\providecommand{\newblock}{\relax}
\providecommand{\bibinfo}[2]{#2}
\providecommand{\BIBentrySTDinterwordspacing}{\spaceskip=0pt\relax}
\providecommand{\BIBentryALTinterwordstretchfactor}{4}
\providecommand{\BIBentryALTinterwordspacing}{\spaceskip=\fontdimen2\font plus
\BIBentryALTinterwordstretchfactor\fontdimen3\font minus
  \fontdimen4\font\relax}
\providecommand{\BIBforeignlanguage}[2]{{%
\expandafter\ifx\csname l@#1\endcsname\relax
\typeout{** WARNING: IEEEtranN.bst: No hyphenation pattern has been}%
\typeout{** loaded for the language `#1'. Using the pattern for}%
\typeout{** the default language instead.}%
\else
\language=\csname l@#1\endcsname
\fi
#2}}
\providecommand{\BIBdecl}{\relax}
\BIBdecl

\bibitem[Gousios et~al.(2014)Gousios, Pinzger, and
  Deursen]{gousios2014exploratory}
G.~Gousios, M.~Pinzger, and A.~v. Deursen, ``An exploratory study of the
  pull-based software development model,'' in \emph{Proceedings of the 36th
  International Conference on Software Engineering}, 2014, pp. 345--355.

\bibitem[Barr et~al.(2012)Barr, Bird, Rigby, Hindle, German, and
  Devanbu]{barr2012cohesive}
E.~T. Barr, C.~Bird, P.~C. Rigby, A.~Hindle, D.~M. German, and P.~Devanbu,
  ``Cohesive and isolated development with branches,'' in \emph{International
  Conference on Fundamental Approaches to Software Engineering}.\hskip 1em plus
  0.5em minus 0.4em\relax Springer, 2012, pp. 316--331.

\bibitem[Gousios et~al.(2015)Gousios, Zaidman, Storey, and
  Deursen]{Gousios:2015}
G.~Gousios, A.~Zaidman, M.-A. Storey, and A.~v. Deursen, ``Work practices and
  challenges in pull-based development: The integrator's perspective,'' in
  \emph{2015 IEEE/ACM 37th IEEE International Conference on Software
  Engineering}, vol.~1, 2015, pp. 358--368.

\bibitem[Wan et~al.(2018)Wan, Xia, Hassan, Lo, Yin, and
  Yang]{wan2018perceptions}
Z.~Wan, X.~Xia, A.~E. Hassan, D.~Lo, J.~Yin, and X.~Yang, ``Perceptions,
  expectations, and challenges in defect prediction,'' \emph{IEEE Transactions
  on Software Engineering}, vol.~46, no.~11, pp. 1241--1266, 2018.

\bibitem[Kamei and Shihab(2016)]{kamei2016defect}
Y.~Kamei and E.~Shihab, ``Defect prediction: Accomplishments and future
  challenges,'' in \emph{2016 IEEE 23rd international conference on software
  analysis, evolution, and reengineering (SANER)}, vol.~5.\hskip 1em plus 0.5em
  minus 0.4em\relax IEEE, 2016, pp. 33--45.

\bibitem[Saliu and Ruhe(2005)]{saliu2005supporting}
O.~Saliu and G.~Ruhe, ``Supporting software release planning decisions for
  evolving systems,'' in \emph{29th Annual IEEE/NASA Software Engineering
  Workshop}.\hskip 1em plus 0.5em minus 0.4em\relax IEEE, 2005, pp. 14--26.

\bibitem[Adams and McIntosh(2016)]{adams2016modern}
B.~Adams and S.~McIntosh, ``Modern release engineering in a nutshell--why
  researchers should care,'' in \emph{2016 IEEE 23rd international conference
  on software analysis, evolution, and reengineering (SANER)}, vol.~5.\hskip
  1em plus 0.5em minus 0.4em\relax IEEE, 2016, pp. 78--90.

\bibitem[Rodr{\'\i}guez-P{\'e}rez et~al.(2018)Rodr{\'\i}guez-P{\'e}rez, Robles,
  and Gonz{\'a}lez-Barahona]{rodriguez2018reproducibility}
G.~Rodr{\'\i}guez-P{\'e}rez, G.~Robles, and J.~M. Gonz{\'a}lez-Barahona,
  ``Reproducibility and credibility in empirical software engineering: A case
  study based on a systematic literature review of the use of the szz
  algorithm,'' \emph{Information and Software Technology}, vol.~99, pp.
  164--176, 2018.

\bibitem[{\'S}liwerski et~al.(2005){\'S}liwerski, Zimmermann, and
  Zeller]{sliwerski2005changes}
J.~{\'S}liwerski, T.~Zimmermann, and A.~Zeller, ``When do changes induce
  fixes?'' \emph{ACM sigsoft software engineering notes}, vol.~30, no.~4, pp.
  1--5, 2005.

\bibitem[Kim et~al.(2006)Kim, Zimmermann, Pan, James~Jr,
  et~al.]{kim2006automatic}
S.~Kim, T.~Zimmermann, K.~Pan, E.~James~Jr \emph{et~al.}, ``Automatic
  identification of bug-introducing changes,'' in \emph{21st IEEE/ACM
  international conference on automated software engineering (ASE'06)}.\hskip
  1em plus 0.5em minus 0.4em\relax IEEE, 2006, pp. 81--90.

\bibitem[Davies et~al.(2014)Davies, Roper, and Wood]{davies2014comparing}
S.~Davies, M.~Roper, and M.~Wood, ``Comparing text-based and dependence-based
  approaches for determining the origins of bugs,'' \emph{Journal of Software:
  Evolution and Process}, vol.~26, no.~1, pp. 107--139, 2014.

\bibitem[Neto et~al.(2018)Neto, da~Costa, and Kulesza]{neto2018impact}
E.~C. Neto, D.~A. da~Costa, and U.~Kulesza, ``The impact of refactoring changes
  on the szz algorithm: An empirical study,'' in \emph{2018 IEEE 25th
  International Conference on Software Analysis, Evolution and Reengineering
  (SANER)}.\hskip 1em plus 0.5em minus 0.4em\relax IEEE, 2018, pp. 380--390.

\bibitem[{Neto} et~al.(2019){Neto}, d.~{Costa}, and {Kulesza}]{neto2019}
E.~C. {Neto}, D.~A. d.~{Costa}, and U.~{Kulesza}, ``Revisiting and improving
  szz implementations,'' in \emph{2019 ACM/IEEE International Symposium on
  Empirical Software Engineering and Measurement (ESEM)}, 2019, pp. 1--12.

\bibitem[Rosa et~al.(2021)Rosa, Pascarella, Scalabrino, Tufano, Bavota, Lanza,
  and Oliveto]{pascarella2021evaluating}
G.~Rosa, L.~Pascarella, S.~Scalabrino, R.~Tufano, G.~Bavota, M.~Lanza, and
  R.~Oliveto, ``Evaluating {SZZ} implementations through a developer-informed
  oracle,'' in \emph{2021 IEEE/ACM 43rd International Conference on Software
  Engineering (ICSE)}.\hskip 1em plus 0.5em minus 0.4em\relax IEEE, 2021, pp.
  436--447.

\bibitem[Da~Costa et~al.(2016)Da~Costa, McIntosh, Shang, Kulesza, Coelho, and
  Hassan]{da2016framework}
D.~A. Da~Costa, S.~McIntosh, W.~Shang, U.~Kulesza, R.~Coelho, and A.~E. Hassan,
  ``A framework for evaluating the results of the szz approach for identifying
  bug-introducing changes,'' \emph{IEEE Transactions on Software Engineering},
  vol.~43, no.~7, pp. 641--657, 2016.

\bibitem[Lenarduzzi et~al.(2020)Lenarduzzi, Palomba, Taibi, and
  Tamburri]{Lenarduzzi:2020}
V.~Lenarduzzi, F.~Palomba, D.~Taibi, and D.~A. Tamburri, ``Openszz: A free,
  open-source, web-accessible implementation of the szz algorithm,'' ser. ICPC
  '20, 2020, pp. 446--450.

\bibitem[Borg et~al.(2019)Borg, Svensson, Berg, and Hansson]{borg2019szz}
M.~Borg, O.~Svensson, K.~Berg, and D.~Hansson, ``Szz unleashed: an open
  implementation of the szz algorithm-featuring example usage in a study of
  just-in-time bug prediction for the jenkins project,'' in \emph{Proceedings
  of the 3rd ACM SIGSOFT International Workshop on Machine Learning Techniques
  for Software Quality Evaluation}, 2019, pp. 7--12.

\bibitem[Spadini et~al.(2018)Spadini, Aniche, and
  Bacchelli]{spadini2018pydriller}
\BIBentryALTinterwordspacing
D.~Spadini, M.~Aniche, and A.~Bacchelli, ``{PyDriller: Python framework for
  mining software repositories},'' in \emph{Proceedings of the 2018 26th ACM
  Joint Meeting on European Software Engineering Conference and Symposium on
  the Foundations of Software Engineering - ESEC/FSE 2018}.\hskip 1em plus
  0.5em minus 0.4em\relax New York, New York, USA: ACM Press, 2018, pp.
  908--911. [Online]. Available:
  \url{http://dl.acm.org/citation.cfm?doid=3236024.3264598}
\BIBentrySTDinterwordspacing

\bibitem[Herbold et~al.(2019)Herbold, Trautsch, Trautsch, and
  Ledel]{herbold2019issues}
S.~Herbold, A.~Trautsch, F.~Trautsch, and B.~Ledel, ``Issues with szz: An
  empirical assessment of the state of practice of defect prediction data
  collection,'' \emph{arXiv preprint arXiv:1911.08938}, 2019.

\bibitem[T{\'o}th et~al.(2016)T{\'o}th, Gyimesi, and Ferenc]{Toth:2016}
Z.~T{\'o}th, P.~Gyimesi, and R.~Ferenc, ``A public bug database of github
  projects and its application in bug prediction,'' in \emph{Computational
  Science and Its Applications -- ICCSA 2016}, O.~Gervasi, B.~Murgante,
  S.~Misra, A.~M.~A. Rocha, C.~M. Torre, D.~Taniar, B.~O. Apduhan, E.~Stankova,
  and S.~Wang, Eds.\hskip 1em plus 0.5em minus 0.4em\relax Cham: Springer
  International Publishing, 2016, pp. 625--638.

\bibitem[Pascarella et~al.(2019)Pascarella, Palomba, and
  Bacchelli]{pascarella2019}
L.~Pascarella, F.~Palomba, and A.~Bacchelli, ``Fine-grained just-in-time defect
  prediction,'' \emph{Journal of Systems and Software}, vol. 150, pp. 22--36,
  2019.

\bibitem[Kim et~al.(2008)Kim, Whitehead, and Zhang]{Kim:2008}
S.~Kim, E.~J. Whitehead, and Y.~Zhang, ``Classifying software changes: Clean or
  buggy?'' \emph{IEEE Transactions on Software Engineering}, vol.~34, no.~2,
  pp. 181--196, 2008.

\bibitem[Wen et~al.(2016)Wen, Wu, and Cheung]{Wen:2016}
M.~Wen, R.~Wu, and S.-C. Cheung, ``Locus: Locating bugs from software
  changes,'' in \emph{2016 31st IEEE/ACM International Conference on Automated
  Software Engineering (ASE)}, 2016, pp. 262--273.

\bibitem[{Yan} et~al.(2020){Yan}, {Xia}, {Fan}, {Hassan}, {Lo}, and
  {Li}]{yan2020}
M.~{Yan}, X.~{Xia}, Y.~{Fan}, A.~E. {Hassan}, D.~{Lo}, and S.~{Li},
  ``Just-in-time defect identification and localization: A two-phase
  framework,'' \emph{IEEE Transactions on Software Engineering}, pp. 1--1,
  2020.

\bibitem[Kim et~al.(2007)Kim, Zimmermann, Whitehead~Jr, and
  Zeller]{kim2007predicting}
S.~Kim, T.~Zimmermann, E.~J. Whitehead~Jr, and A.~Zeller, ``Predicting faults
  from cached history,'' in \emph{29th International Conference on Software
  Engineering (ICSE'07)}.\hskip 1em plus 0.5em minus 0.4em\relax IEEE, 2007,
  pp. 489--498.

\bibitem[Rahman et~al.(2011{\natexlab{a}})Rahman, Posnett, Hindle, Barr, and
  Devanbu]{Rahman:2011}
F.~Rahman, D.~Posnett, A.~Hindle, E.~Barr, and P.~Devanbu, ``Bugcache for
  inspections: Hit or miss?'' in \emph{Proceedings of the 19th ACM SIGSOFT
  Symposium and the 13th European Conference on Foundations of Software
  Engineering}, ser. ESEC/FSE '11, 2011, pp. 322--331.

\bibitem[Chen and Jiang(2019)]{chen2019extracting}
B.~Chen and Z.~M.~J. Jiang, ``Extracting and studying the
  logging-code-issue-introducing changes in java-based large-scale open source
  software systems,'' \emph{Empirical Software Engineering}, vol.~24, no.~4,
  pp. 2285--2322, 2019.

\bibitem[Fan et~al.(2019)Fan, Xia, da~Costa, Lo, Hassan, and Li]{Fan:2019}
Y.~Fan, X.~Xia, D.~A. da~Costa, D.~Lo, A.~E. Hassan, and S.~Li, ``The impact of
  changes mislabeled by szz on just-in-time defect prediction,'' \emph{IEEE
  Transactions on Software Engineering}, no.~01, pp. 1--1, jul 2019.

\bibitem[Aman et~al.(2019)Aman, Amasaki, Yokogawa, and
  Kawahara]{aman2019empirical}
H.~Aman, S.~Amasaki, T.~Yokogawa, and M.~Kawahara, ``Empirical study of fault
  introduction focusing on the similarity among local variable names.'' in
  \emph{QuASoQ@ APSEC}, 2019, pp. 3--11.

\bibitem[Eyolfson et~al.(2014)Eyolfson, Tan, and Lam]{eyolfson2014correlations}
J.~Eyolfson, L.~Tan, and P.~Lam, ``Correlations between bugginess and
  time-based commit characteristics,'' \emph{Empirical Software Engineering},
  vol.~19, no.~4, pp. 1009--1039, 2014.

\bibitem[Bernardi et~al.(2018)Bernardi, Canfora, Di~Lucca, Di~Penta, and
  Distante]{bernardi2018relation}
M.~L. Bernardi, G.~Canfora, G.~A. Di~Lucca, M.~Di~Penta, and D.~Distante, ``The
  relation between developers' communication and fix-inducing changes: An
  empirical study,'' \emph{Journal of Systems and Software}, vol. 140, pp.
  111--125, 2018.

\bibitem[Izquierdo-Cort{\'a}zar et~al.("2012")Izquierdo-Cort{\'a}zar, Robles,
  and Gonz{\'a}lez-Barahona]{Izquierdo-Cortazar2012}
D.~Izquierdo-Cort{\'a}zar, G.~Robles, and J.~M. Gonz{\'a}lez-Barahona, ``"do
  more experienced developers introduce fewer bugs?",'' in \emph{"Open Source
  Systems: Long-Term Sustainability"}, I.~"Hammouda, B.~Lundell, T.~Mikkonen,
  and W.~Scacchi, Eds.\hskip 1em plus 0.5em minus 0.4em\relax "Berlin,
  Heidelberg": "Springer Berlin Heidelberg", "2012", pp. "268--273".

\bibitem[Tufano et~al.(2017)Tufano, Bavota, Poshyvanyk, Di~Penta, Oliveto, and
  De~Lucia]{tufano2017empirical}
M.~Tufano, G.~Bavota, D.~Poshyvanyk, M.~Di~Penta, R.~Oliveto, and A.~De~Lucia,
  ``An empirical study on developer-related factors characterizing fix-inducing
  commits,'' \emph{Journal of Software: Evolution and Process}, vol.~29, no.~1,
  p. e1797, 2017.

\bibitem[Kalliamvakou et~al.(2014)Kalliamvakou, Gousios, Blincoe, Singer,
  German, and Damian]{kalliamvakou2014promises}
E.~Kalliamvakou, G.~Gousios, K.~Blincoe, L.~Singer, D.~M. German, and
  D.~Damian, ``The promises and perils of mining github,'' in \emph{Proceedings
  of the 11th working conference on mining software repositories}, 2014, pp.
  92--101.

\bibitem[Gousios et~al.(2016)Gousios, Storey, and Bacchelli]{gousios2016work}
G.~Gousios, M.-A. Storey, and A.~Bacchelli, ``Work practices and challenges in
  pull-based development: the contributor's perspective,'' in \emph{2016
  IEEE/ACM 38th International Conference on Software Engineering (ICSE)}.\hskip
  1em plus 0.5em minus 0.4em\relax IEEE, 2016, pp. 285--296.

\bibitem[Zampetti et~al.(2019)Zampetti, Bavota, Canfora, and
  Di~Penta]{zampetti2019study}
F.~Zampetti, G.~Bavota, G.~Canfora, and M.~Di~Penta, ``A study on the interplay
  between pull request review and continuous integration builds,'' in
  \emph{2019 IEEE 26th International Conference on Software Analysis, Evolution
  and Reengineering (SANER)}.\hskip 1em plus 0.5em minus 0.4em\relax IEEE,
  2019, pp. 38--48.

\bibitem[Castelluccio et~al.(2019)Castelluccio, An, and
  Khomh]{castelluccio2019empirical}
M.~Castelluccio, L.~An, and F.~Khomh, ``An empirical study of patch uplift in
  rapid release development pipelines,'' \emph{Empirical Software Engineering},
  vol.~24, no.~5, pp. 3008--3044, 2019.

\bibitem[Dunsmore et~al.(2003)Dunsmore, Roper, and Wood]{dunsmore2003practical}
A.~Dunsmore, M.~Roper, and M.~Wood, ``Practical code inspection techniques for
  object-oriented systems: an experimental comparison,'' \emph{IEEE software},
  vol.~20, no.~4, pp. 21--29, 2003.

\bibitem[Ram et~al.(2018)Ram, Sawant, Castelluccio, and
  Bacchelli]{ram2018makes}
A.~Ram, A.~A. Sawant, M.~Castelluccio, and A.~Bacchelli, ``What makes a code
  change easier to review: an empirical investigation on code change
  reviewability,'' in \emph{Proceedings of the 2018 26th ACM Joint Meeting on
  European Software Engineering Conference and Symposium on the Foundations of
  Software Engineering}, 2018, pp. 201--212.

\bibitem[GitHub()]{GraphQL}
\BIBentryALTinterwordspacing
GitHub. Github graphql api. [Online]. Available:
  \url{https://docs.github.com/en/graphql}
\BIBentrySTDinterwordspacing

\bibitem[Rezk et~al.(2021)Rezk, Kamei, and Mcintosh]{rezk2021ghost}
C.~Rezk, Y.~Kamei, and S.~Mcintosh, ``The ghost commit problem when identifying
  fix-inducing changes: An empirical study of apache projects,'' \emph{IEEE
  Transactions on Software Engineering}, 2021.

\bibitem[Rodr{\'\i}guez-P{\'e}rez et~al.(2020)Rodr{\'\i}guez-P{\'e}rez, Robles,
  Serebrenik, Zaidman, Germ{\'a}n, and Gonzalez-Barahona]{rodriguez2020bugs}
G.~Rodr{\'\i}guez-P{\'e}rez, G.~Robles, A.~Serebrenik, A.~Zaidman, D.~M.
  Germ{\'a}n, and J.~M. Gonzalez-Barahona, ``How bugs are born: a model to
  identify how bugs are introduced in software components,'' \emph{Empirical
  Software Engineering}, vol.~25, no.~2, pp. 1294--1340, 2020.

\bibitem[Mozilla()]{bugbug}
\BIBentryALTinterwordspacing
Mozilla. Bugbug open source ml project. [Online]. Available:
  \url{https://zenodo.org/record/4911346}
\BIBentrySTDinterwordspacing

\bibitem[Manning et~al.(2010)Manning, Raghavan, and
  Sch{\"u}tze]{manning2010introduction}
C.~Manning, P.~Raghavan, and H.~Sch{\"u}tze, ``Introduction to information
  retrieval,'' \emph{Natural Language Engineering}, vol.~16, no.~1, pp.
  100--103, 2010.

\bibitem[Rahman et~al.(2011{\natexlab{b}})Rahman, Posnett, Hindle, Barr, and
  Devanbu]{rahman2011bugcache}
F.~Rahman, D.~Posnett, A.~Hindle, E.~Barr, and P.~Devanbu, ``Bugcache for
  inspections: hit or miss?'' in \emph{Proceedings of the 19th ACM SIGSOFT
  symposium and the 13th European conference on Foundations of software
  engineering}, 2011, pp. 322--331.

\bibitem[Zhang and Tsai(2003)]{zhang2003machine}
D.~Zhang and J.~J. Tsai, ``Machine learning and software engineering,''
  \emph{Software Quality Journal}, vol.~11, no.~2, pp. 87--119, 2003.

\bibitem[Chawla et~al.(2002)Chawla, Bowyer, Hall, and
  Kegelmeyer]{chawla2002smote}
N.~V. Chawla, K.~W. Bowyer, L.~O. Hall, and W.~P. Kegelmeyer, ``Smote:
  synthetic minority over-sampling technique,'' \emph{Journal of artificial
  intelligence research}, vol.~16, pp. 321--357, 2002.

\bibitem[Ling and Sheng(2008)]{ling2008cost}
C.~X. Ling and V.~S. Sheng, ``Cost-sensitive learning and the class imbalance
  problem,'' \emph{Encyclopedia of machine learning}, vol. 2011, pp. 231--235,
  2008.

\bibitem[Abdi and Williams(2010)]{abdi2010principal}
H.~Abdi and L.~J. Williams, ``Principal component analysis,'' \emph{Wiley
  interdisciplinary reviews: computational statistics}, vol.~2, no.~4, pp.
  433--459, 2010.

\bibitem[Saito and Rehmsmeier(2015)]{saito2015precision}
T.~Saito and M.~Rehmsmeier, ``The precision-recall plot is more informative
  than the roc plot when evaluating binary classifiers on imbalanced
  datasets,'' \emph{PloS one}, vol.~10, no.~3, p. e0118432, 2015.

\end{thebibliography}

	\begin{IEEEbiography}[{\includegraphics[width=1in,height=1.25in,clip,keepaspectratio]{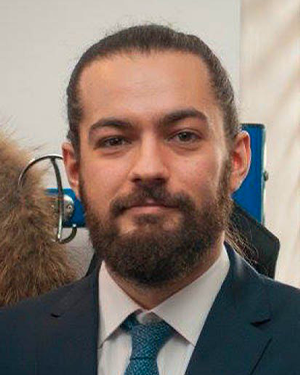}}]{Fernando Petrulio}
		is an Italian PhD Student in Software Engineering at the University of Zurich and a member of ZEST since May 2019.

		He obtained both Bachelor's degree in Computer Science (December 2016) and Master's degree in Computer Science with a Specialization in Data Science (February 2019) at Università degli Studi di Salerno. He spent a period in UZH with the International Mobility Program, during which he produced my Thesis Work under the supervision of Prof. Bacchelli. His main interests are in defect prediction and empirical data analysis.
	\end{IEEEbiography}

	\begin{IEEEbiography}[{\includegraphics[width=1in,height=1.25in,clip,keepaspectratio]{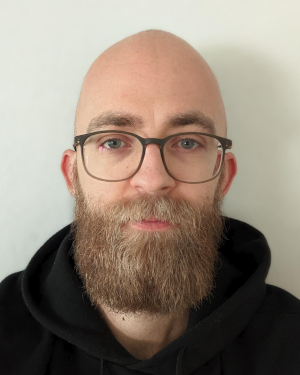}}]{David Ackermann}
		
		is a Software Engineer that received his M.Sc. in Data Science and B.Sc. in Business Informatics from the University of Zurich. His interests are centered on making fuzz testing more effective by combining it with different machine learning algorithms and leveraging the human-in-the-loop.
	\end{IEEEbiography}
	
	\begin{IEEEbiography}[{\includegraphics[width=1in,height=1.25in,clip,keepaspectratio]{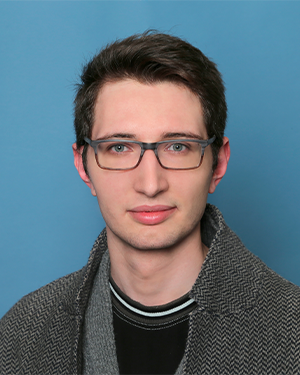}}]{Enrico Fregnan}
		is a Ph.D. student in the Zurich Empirical Software engineering Team (ZEST) at the University of Zurich. He received his bachelor’s degree at Politecnico di Milano, Italy and his master’s degree at Delft University of Technology, The Netherlands. His research focuses on investigating how to support developers during code review.
	\end{IEEEbiography}

	\begin{IEEEbiography}[{\includegraphics[width=1in,height=1.25in,clip,keepaspectratio]{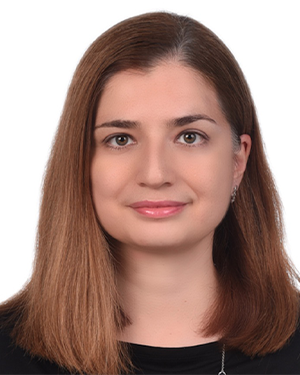}}]{G\"{u}l Calikli}
	received Ph.D. and master’s degrees in Computer Engineering and bachelor’s degree in Mechanical Engineering from Boğazici University in Istanbul, Turkey. She is a lecturer (assistant professor) in Software Engineering with the School of Computing Science at the University of Glasgow, United Kingdom.
	\end{IEEEbiography}
	
	\begin{IEEEbiography}[{\includegraphics[width=1in,height=1.25in,clip,keepaspectratio]{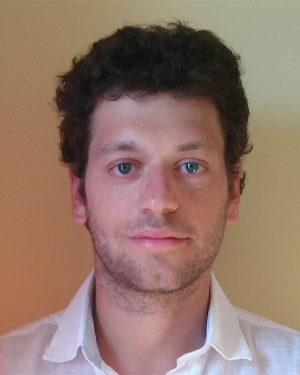}}]{Marco Castelluccio}
	is an engineering manager at Mozilla. He received his Bachelor, his Master and his Ph.D. of Computer Science Engineering at University of Napoli Federico II. His research interests include software engineering, software reliability, software maintenance and evolution, software analytics, and artificial intelligence.
	\end{IEEEbiography}
	
	\begin{IEEEbiography}[{\includegraphics[width=1in,height=1.25in,clip,keepaspectratio]{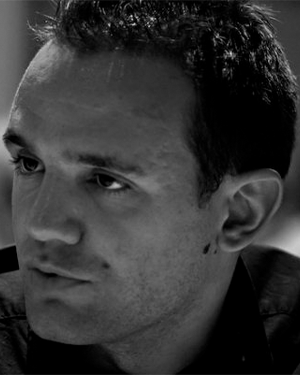}}]{Sylvestre Ledru}
		is a director of engineering at Mozilla and a volunteer on Debian, LLVM and Rust, with more than 20 years of experience in the software development industry. He is focusing his efforts on low level systems, software development processes, code quality and security.
	\end{IEEEbiography}

	\begin{IEEEbiography}[{\includegraphics[width=1in,height=1.25in,clip,keepaspectratio]{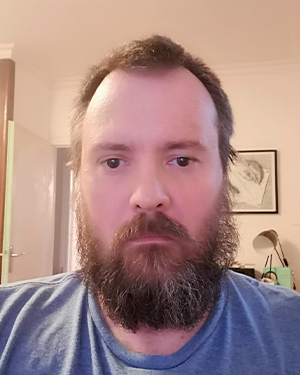}}]{Calixte Denizet}
		 is a senior software engineer at Mozilla and a former math teacher. He received his Master of Mathematics at the University of Caen in 1997.
	\end{IEEEbiography}

	\begin{IEEEbiography}[{\includegraphics[width=1in,height=1.25in,clip,keepaspectratio]{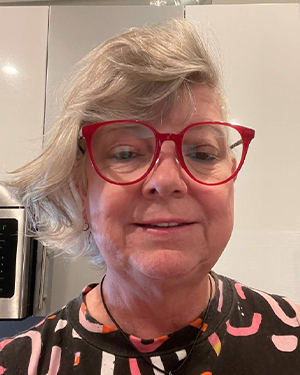}}]{Emma Humpries}
		 works as a developer experience engineer at Bandcamp. They received their Master of Science in Economics from the University of Wisconsin, Madison.
	\end{IEEEbiography}
	
	\begin{IEEEbiography}[{\includegraphics[width=1in,height=1.25in,clip,keepaspectratio]{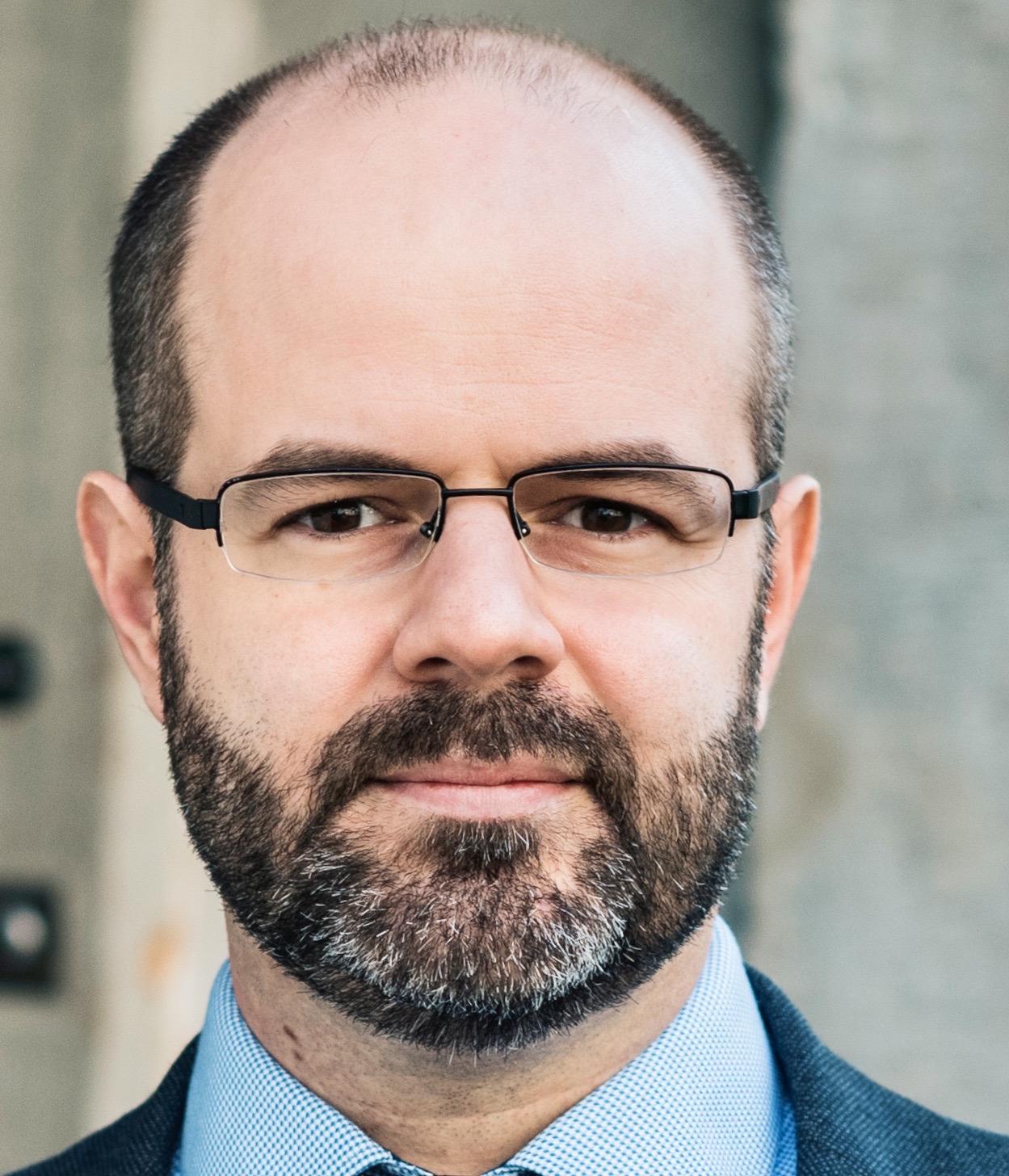}}]{Alberto Bacchelli}
		received the bachelor’s and master’s degrees in computer science from the University of Bologna, Italy, and the PhD degree in software engineering from the Università della Svizzera Italiana, Switzerland. He is an associate professor of Empirical Software Engineering with the Department of Informatics in the Faculty of Business, Economics and Informatics at the University of Zurich, Switzerland.
	\end{IEEEbiography}

\end{document}